\documentclass[12pt]{article} 
\usepackage{epsf,amsmath}

\begin{document}
\setlength{\parskip}{2ex} 
\setlength{\parindent}{0em}
\setlength{\baselineskip}{3ex} 
\renewcommand{\theequation}{\thesection.\arabic{equation}}
  \newcommand{\onefigure}[2]{\begin{figure}[htbp]
         \caption{\small #2\label{#1}(#1)}
         \end{figure}}
\renewcommand{\onefigure}[2]{\begin{figure}[htbp]
        \begin{center}\leavevmode\epsfbox{#1.eps}\end{center}
         \caption{\small #2\label{#1}}
         \end{figure}}
\newcommand{\comment}[1]{} 
\newcommand{\myref}[1]{(\ref{#1})}
\newcommand{\secref}[1]{sec.~\protect\ref{#1}}
\newcommand{\figref}[1]{Fig.~\protect\ref{#1}}
\newcommand{\mathbold}[1]{\mbox{\boldmath $\bf#1$}}
\newcommand{\mJ}{\mathbold{J}}
\newcommand{\momega}{\mathbold{\omega}}
\newcommand{\bz}{{\bf z}}
\def\bbbz{{\sf Z\!\!\!Z}}
\newcommand{\PP}{\mbox{I}\!\mbox{P}}
\def\sl2z{SL(2,\bbbz)}
\newcommand{\bbbq}{I\!\!Q}
\newcommand{\be}{\begin{equation}}
\newcommand{\ee}{\end{equation}}
\newcommand{\bea}{\begin{eqnarray}}
\newcommand{\eea}{\end{eqnarray}}
\newcommand{\nn}{\nonumber}
\newcommand{\unit}{1\!\!1}
\newcommand{\half}{\frac{1}{2}}
\newcommand{\shalf}{\mbox{$\half$}}
\newcommand{\transform}[1]{
   \stackrel{#1}{-\hspace{-1.2ex}-\hspace{-1.2ex}\longrightarrow}}
\newcommand{\inter}[2]{\null^{\#}(#1\cdot#2)}
\newcommand{\lprod}[2]{\vec{#1}\cdot\vec{#2}}
\newcommand{\mult}[1]{{\cal N}(#1)}
\newcommand{\Bn}{{\cal B}_N}
\newcommand{\B}{{\cal B}}
\newcommand{\Beight}{{\cal B}_8}
\newcommand{\Bnine}{{\cal B}_9}
\newcommand{\Eman}{\widehat{\cal E}_N}
\newcommand{\C}{{\cal C}}
\newcommand{\Q}{Q\!\!\!Q}
\newcommand{\comp}{C\!\!\!C}
\newcommand{\D}{{\cal D}}
\newcommand{\mA}{{\mathbold A}}
\def\malpha{\mathbold{\alpha}}

\noindent
\thispagestyle{empty}
{\flushright{\small AEI-1999-36\\ MIT-CTP-2927\\hep-th/0002127\\}}

\vspace{.3in}
\begin{center}\Large {\bf Duality and Weyl Symmetry of  7-brane
Configurations}

\end{center}

\vspace{.3in}
\begin{center}
{\large Tam\'{a}s Hauer}
\vspace{.1in}

Max-Planck-Institut f\"ur Gravitationsphysik,\\
Albert-Einstein-Institut, D-14476 Golm, Germany.
\vspace{.2in}

{\large Amer Iqbal  \,and\,  Barton Zwiebach}
\vspace{.1in}

Center for Theoretical Physics,\\
Massachusetts Institute of Technology,\\
Cambridge, Massachusetts 02139, U.S.A.
\vspace{.2in}

E-mail: {\tt hauer@aei-potsdam.mpg.de, amer@mit.edu,
zwiebach@mitlns.mit.edu}
\end{center}
\begin{center} February 2000
\end{center}

\vspace{0.1in}

\begin{abstract}
Extending earlier results on the duality symmetries of three-brane
probe theories we define the duality subgroup of $\sl2z$ as the
symmetry group of the background 7-branes configurations.  
We establish that the action of Weyl reflections is 
implemented on junctions by  brane
transpositions that amount to exchanging branes
that can be connected by open strings. This enables us to 
characterize duality groups of brane configurations
by  a map to the symmetry group of the
Dynkin diagram.
We compute the duality groups and their actions for
all localizable 7-brane configurations.
Surprisingly, for the case of affine configurations
there are brane transpositions leaving them invariant
but acting nontrivially on the charges 
of junctions.


\end{abstract}
\newpage

\section{Introduction}
\setcounter{equation}{0}

The four dimensional ${\cal N}\!=\!2$ supersymmetric $SU(2)$
Yang-Mills theory with $N_f=4$ is a theory with full $\sl2z$ duality
symmetry and global $so(8)$ symmetry \cite{seibergwitten}. The duality
symmetry has nontrivial implications for the dyonic spectrum of the
theory. In fact the dyons fall into the ${\bf 8_v}, {\bf 8_s}$ or
${\bf 8_c}$ representations according to their dyonic charges $(p,q)$.
In addition, duality transformations permute such representations.
This action can be characterized nicely as a homomorphism from the
duality group $\sl2z$ to the permutation group on three objects, the
group of graph automorphisms of the $so(8)$ Dynkin diagram.

This quantum field theory can be viewed as living on a D3 brane probe
of a 7-brane background \cite{sen,rutgers}. The 7-brane background
includes the six 7-branes that compose the $D_4$ singularity of the
Kodaira classification. Both the duality symmetry of the theory, and
its interaction with the $D_4$ symmetry carried by the branes can be
understood simply in terms of the 7-brane background
\cite{imamura,DZ,DHIZ,nekrasov,ansar}.  It is the purpose of the
present paper to give a precise definition of the duality group
associated to an {\it arbitrary} 7-brane configuration, and a
description of its interaction with the Lie algebra carried by the
configuration.  Partial results were given in \cite{DHIZ} where the
duality group of a brane configuration with monodromy $K \in \sl2z$
was thought to be simply the stabilizer subgroup of $K$ in $\sl2z$.
While correct for many cases, this is not always true and further
conditions must be satisfied.

The duality symmetry of the four-dimensional effective theory is the
remnant of the $\sl2z$ symmetry of IIB string theory. Consider the
simplest case of a D3 in the vicinity of D7-branes. The three-brane is
$\sl2z$-symmetric while the D7-branes are left invariant by the
subgroup generated by $T\in\sl2z$; therefore the effective theory on
the D3 still carries this subgroup as a duality symmetry (a trivial
one because the theory does not have magnetic states).  When 7-branes
of different charges are involved only $\unit$ and $-\unit\in\sl2z$
leave each brane invariant. Those nontrivial transformations, however,
which map the charges of the 7-branes to each other, do act as a
duality symmetry of the effective theory because the transformed
configuration is indistinguishable from the original one (although the
position in the moduli space changes in general).  The subtle task of
identifying those $\sl2z$ matrices which permute the charges of a
given 7-brane configuration will be one subject of our present work.
The difficulty arises because there is a large redundancy in
characterizing a background by a list of 7-brane charges. To this end
we will define the equivalence classes of configurations along the
lines of \cite{mgthbz,finite}: equivalent 7-brane setups are related
to each other by the process of moving branch cuts of 7-branes through
each other. Then we look for $\sl2z$ transformations which map a given
7-brane configuration to another one in its equivalence class.

Having found the ``unbroken part'' of $\sl2z$ one should identify how
this duality symmetry acts on the spectrum of the theory. As is well
known from the Seiberg-Witten example, the duality group acts through
the automorphism group of the root lattice of the algebra carried by
the 7-branes.  Such automorphisms are of two kinds: those related to
the automorphisms of the Dynkin diagram, and those arising from Weyl
reflections. In elucidating the general theory we have found it
necessary to identify how Weyl symmetries are represented on the
7-brane configurations.

The states of D3 brane probe theories are strings or string junctions
stretched between the D3 and (some of the) background 7-branes. When
the 7-branes are on top of each other, the 7+1 dimensional low energy
model is a Yang-Mills theory with some gauge algebra $\cal G$ while
the spectrum of the 4-dimensional D3-theory furnishes a representation
of $\cal G$ which is a global symmetry. When the 7-branes do not
coincide (and many times they cannot) the states are no longer
degenerate but $\cal G$ is still a useful spectrum generating algebra.
In particular states constituting the orbits of the Weyl group of
$\cal G$ are of the same multiplicity. Indeed, Weyl reflections derive
from an ambiguity in the choice of a base of simple roots, and states
related by Weyl reflections are really physically equivalent.

\onefigure{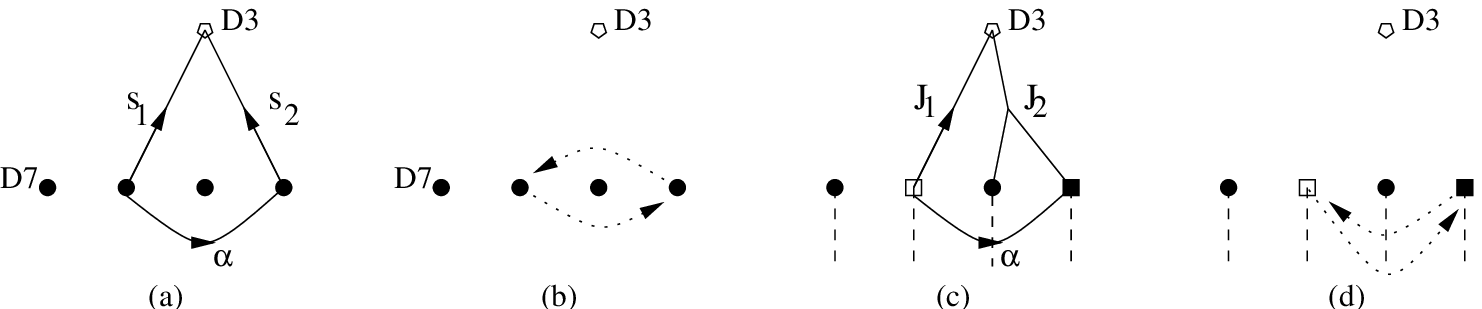}{Weyl transformations correspond to branch cut moves
  of the 7-brane configuration.}

How do Weyl transformations act on junctions?  Consider a
configuration of D7-branes.  \figref{sun}(a) shows two strings, ${\bf
  s}_1$ and ${\bf s}_2$ which are in the fundamental representation of
$su(n)$ and are mapped to each other by the Weyl reflection with the
root corresponding to $\alpha$.  This is the case because using
familiar intersection rules and composition of open strings we see
that ${\bf s}_1 + ({\bf s}_1 \cdot \malpha) \malpha = {\bf s}_1
-\malpha = {\bf s}_2$.  Physically, this Weyl reflection is
implemented by the exchange process of the two D7-branes (see
\figref{sun}(b)), giving us an identical theory in which the two
states ${\bf s}_1$ and ${\bf s}_2$ are replaced with each other.  The
cases involving mutually nonlocal 7-branes like the one in
\figref{sun}(c) are treated similarly except that one should be
careful when exchanging 7-branes because the path along which they are
moved is relevant.  Anticipating a more extended discussion later in
the paper, we note that the rules for transpositions of 7-branes imply
that the transformation shown in \figref{sun}(d) leaves the brane
configuration invariant.  In the process of transposition the two
junctions $\mJ_1$ and $\mJ_2=\mJ_1 + \malpha\,(\mJ_1\cdot\malpha)$ can
be shown to transform into each other as a consequence of prong
creation taking place when 7-branes move through strings.  Thus
$\mJ_1$ and $\mJ_2$, while not equivalent junctions, represent states
that are physically indistinguishable.  We show generally that Weyl
reflections act on junctions as automorphisms of the junction lattice
generated by exchanging 7-branes through a path along which they are
mutually local.

Weyl and duality transformations are different in nature: the former
is an automorphism of the junction lattice mapping states into states
with the same asymptotic charges, while the latter, acting through an
element of $\sl2z$ can change the asymptotic charges of junctions.  In
addition to this action, duality transformations are characterized by
the kind of transformations they generate on the Lie-algebraic data
characterizing the junctions.  We show that these amount to
automorphisms of the root lattice $Q$ of the Lie algebra carried by
the 7-branes. Sometimes duality transformations give Weyl
transformations of the root lattice, in which case they relate states
with possibly different asymptotic charges, but having essentially
equivalent Lie-algebraic data.  On the other hand, some duality
transformations map to automorphisms that correspond to symmetries of
the Dynkin diagram. Such transformations are never Weyl reflections,
and relate states in different Weyl orbits or different
representations.

We define the duality group $\D$ of a brane configuration
as the subgroup of $\sl2z$ whose effect on the brane
configuration can be undone by crossing transformations.
A 7-brane background carrying a Lie
algebra
${\cal G}$ has a duality group $\D$ that interacts nontrivially with ${\cal G}$ when the
symmetry group $\Gamma$ of the Dynkin diagram of ${\cal G}$ is
nontrivial.
We will see that in general there is no canonical map ${\cal D}\mapsto
\Gamma$  because of the surprising fact that invariant transpositions,
transpositions that leave the brane configuration invariant, do
not always act as Weyl transformations on the Lie algebraic 
data of the junction. Sometimes they may act as some
outer automorphism of the root lattice. Since $\Gamma$ precisely
represents outer automorphisms, the lack of a canonical map
simply results because after an $\sl2z$ duality, the
restoring crossing transformation is ambiguous up to 
invariant transpositions. 

If the action of the set invariant transpositions on
the Lie algebraic data induces a Weyl transformation then the above
map is well defined.  This is the case for all the configurations realizing finite
algebras. 
Particularly interesting are the cases of configurations carrying
affine exceptional symmetries \cite{dewolfe, infinite,
yang,SZ,HI}. In these cases the Dynkin diagrams typically have
nontrivial automorphism groups and therefore dualities interact rather
nontrivially with the Lie algebra data.  
Also for these configurations
not all invariant transpositions induce Weyl transformations on the
Lie algebraic data of a junction. Therefore for these configurations
the above map is only well defined with respect to a particular set of
transpositions used to undo the $\sl2z$ transformations. 
 In this
paper we compute the duality groups $\D$ and the maps to the
symmetry groups $\Gamma$ of Dynkin diagrams for all brane
configurations that can be localized in a compactification of type
IIB string theory (see \cite{SZ}).

 Let us mention two questions that we have not discussed in this
paper.
 While the action of dualities on root junctions determines
fully the
 action of duality on weight junctions for finite algebras,
in the case
 of affine Kac-Moody algebras more work is necessary to
understand how
 dualities act on junctions that represent general
weight vectors.
 Thus, for the affine exceptional configurations
discussed here, our
 results are restricted to junctions in the root
lattice.  Second, we
 have not characterized the duality action for
the configuration giving
 the Lie algebra $\widehat E_9$. Since this is
not an affine Kac-Moody
 algebra the automorphism group of its root
lattice appears to be
 unfamiliar.
 
 In this paper, as in our
previous ones \cite{finite, infinite} we have
 focused on the symmetry
aspects of 7-brane configurations. Another
 line of works dealing
mostly with tests of F-theory/heterotic duality
 has been reviewed in
\cite{lerche} and some recent works of interest
 are \cite{barrozo,
imamura2}.
 
 The paper is arranged as follows. In section 2 we define
equivalent
 7-brane backgrounds and the duality group of a 7-brane
configuration.
 Section 3 introduces the Weyl transformations in
7-brane language. In
 section 4 we discuss the homomorphism between
the duality (effective)
 group and the automorphism group of the
Dynkin diagram of the
 underlying algebra. In sections 5 and 6 we
systematically compute
 duality groups and homomorphisms for 7-brane
configurations
 corresponding to finite and affine algebras,
respectively.
 

\section{Brane configurations, duality and crossing transformations}
\setcounter{equation}{0}

The purpose of this section is to give a precise definition and begin
the characterization of the duality group ${\cal D}(w)$ of a
configuration $w$ of 7-branes.  The duality group is a subgroup of
$\sl2z$ leaving the brane configuration invariant in a sense we
describe in detail. Roughly, an $\sl2z$ element belongs to the duality
group when its action on the branes can be undone by a crossing
transformation, that is successive operations of brane transpositions.
We show that both the monodromy $K(w)$ of a configuration and the
$(-\unit)$ matrix belong to the duality group ${\cal D}(w)$.  We also
examine how crossing transformations act on junctions, and that leads
to the understanding of how dualities act on junctions.  In particular
we show that the duality transformation $K(w)$ can be defined to leave
invariant junctions having zero asymptotic charge.  The general
characterization of duality groups by their action on junctions is
left for the next section.

Our presentation here will be formal for the sake of precision and
brevity.  Certainly the idea of the duality group of a brane
configuration is not new. The duality groups of Seiberg-Witten $N=2$
SYM theories, in particular those of the $N_f \leq 4$ $SU(2)$
theories
have been studied in detail and correspond to the duality groups of
the 7-brane configurations where these four-dimensional theories
arise
on a D3 brane probe. A definition of the duality group for $D_{N>4}$
and $E_6,E_7,E_8$ was given in \cite{DHIZ}. The present one is a
refined version of that definition, applicable to other brane
configurations as well.

We define what we mean by a 7-brane configuration keeping the
canonical picture in mind: the branes are arranged along the real axis
with their branch cuts going downwards and the monodromies are listed
in the order the branes appear from left to right \cite{mgthbz,DZ}.

\subsection{Definitions and properties}

\begin{itemize}
\item{\bf Definition:} {\em 7-brane configuration} or simply a {\em
    configuration} is defined to be a word of $\sl2z$ matrices
  conjugate to $T^{-1}$. That is, a 7-brane is characterized by the
  monodromy of the axion-dilaton field $\tau$ around it. Via
  introduction of branch cuts $\tau$ is made single valued and $[A]$
  stands for a 7-brane with a cut where $\tau$ jumps by $A\in\sl2z$.
  To define a 7-brane background we list the branes with cuts going
  downwards from left to right. The set of all configurations
  consisting of $n$ 7-branes we denote as ${\cal C}_n$:
  \begin{eqnarray}
    {\cal C}_n \ni w &=& [A_1][A_2]\ldots [A_n]
    \hspace{2cm} A_i=g_{i}T^{-1}g_{i}^{-1},\;\; g_i\in\sl2z, \nn\\
    K(w)&\equiv& A_n\ldots A_2A_1\in\sl2z.
  \end{eqnarray}
  $K(w)$ is the overall monodromy associated with the configuration
  $w$.  The purpose of the square bracket is to distinguish words of
  matrices from products of them, i.e. $[A_1A_2]\in{\cal C}_1$
denotes
  a one-letter word of the matrix $A=A_1A_2$ while
$[A_1][A_2]\in{\cal
    C}_2$ is a two letter word made of these two.
\item{\bf Definition:} {\em $\sl2z$ action on ${\cal C}_n$}. $\sl2z$
  symmetry of IIB transforms $\tau$ and as a consequence the monodromy
  of a 7-brane is conjugated by this transformation $g\in\sl2z$. The
  image of a configuration is simply the word of the transformed
  matrices:
  \begin{eqnarray}
    \sl2z\ni g:{\cal C}_n\rightarrow {\cal C}_n\hspace{.5in}
    [A_1][A_2]\ldots [A_n]\mapsto 
    [gA_1g^{-1}][gA_2g^{-1}]\ldots [gA_ng^{-1}].
    \label{sl2zdef}
  \end{eqnarray}
\item{\bf Definition:} {\em Transposition} of 7-branes. The position
  of the branch cuts of the 7-branes are unphysical and they can be
  relocated by performing an immaterial $\sl2z$ transformation on all
  physical parameters in a selected region. In particular the relative
  order of the cuts can be changed but the 7-branes through which a
  branch cut is moved are subject to the same $\sl2z$ transformation.
  The elementary change of a configuration is when a cut of a brane is
  moved through its left hand ($P$) or right hand ($P^{-1}$) neighbor.
  The transposition of the $m$-th and $m+1$-th letter of a word is
  thus performed by the following rule \cite{mgthbz}:
  \begin{eqnarray}
    \begin{array}{lll}
      P_m:&
      \ldots [A_m][A_{m\!+\!1}]\ldots \longmapsto 
      \ldots [A_{m\!+\!1}][A_{m\!+\!1}A_mA_{m\!+\!1}^{-1}]
     \\ &\\
      P_m^{-1}:&
      \ldots [A_m][A_{m\!+\!1}]\ldots \longmapsto 
      \ldots [A_m^{-1}A_{m\!+\!1}A_m][A_m]\ldots . 
    \end{array} 
    \label{transposition}
  \end{eqnarray}
  Notice that $A_m^{-1}A_{m+1}A_m$ is conjugate to $T^{-1}$ if
  $A_{m+1}$ is, therefore the result is indeed in ${\cal C}_n$. Also
  note that in general $P_m^2$ is non trivial.
\item{\bf Property 1:} The transpositions satisfy the following Braid
  group relations:
  \begin{eqnarray}
    P_mP_{m'}=P_{m'}P_m,\;\; \mbox{if}\;\; |m\!-\!m'|>1,
    \hspace{1.5cm}
    P_mP_{m+1}P_m &=& P_{m+1}P_mP_{m+1}.
    \label{braid}
  \end{eqnarray} 
\item{\bf Definition:} The {\em Group of crossing transformations
    $\Bn$} (corresponding to branch cut moves) is defined by its
  action on ${\cal C}_n$. This group is generated by all
  transpositions $\{P_m\}_{m=1}^{n-1}$ subject to the constraint
  \myref{braid}.
\item{\bf Property 2:} The $\sl2z$ transformations as defined in
  \myref{sl2zdef} commute with the transpositions.  If $g$ denotes an
  $\sl2z$ transformation and $b$ a series of transpositions, we have
  $g (b(w)) = b (g(w))$. To prove this it suffices to examine the case
  when $b$ is a single transposition, say $P_1$:
  \begin{eqnarray} 
    \begin{array}{ccccc}
      [A_1][A_2] &&\transform{P_1} && [A_2][A_2A_1A_2^{-1}]
      \\  &&\\ \downarrow g &&&& \downarrow g \\ &&\\
      \null [gA_1g^{-1}][gA_2g^{-1}] && \transform{P_1} && 
      [gA_2g^{-1}][gA_2A_1A_2^{-1}g^{-1}]      
    \end{array}
  \end{eqnarray}
\item{\bf Definition:} {\em Equivalence group.} As explained before,
  if two 7-brane configurations differ by either an overall $\sl2z$ or
  by crossing transformations, they are physically identical. The need
  for distinguishing between configurations up to this equivalence
  motivates the following definition. The {\em equivalence group} of
  ${\cal C}_n$ is the direct product of the two groups: ${\cal E}_n
  \equiv \sl2z \times{\cal B}_n$. The action of elements of ${\cal
    E}_n$ on ${\cal C}_n$ is well-defined due to the commutativity of
  the actions of the two factors, $(g,b) w = g(b(w)) = b(g(w))$.  The
  product on this set is defined by
  $(g_1,b_1)(g_2,b_2)=(g_1g_2,b_1b_2)$ where $g_i\in\sl2z, b_i\in{\cal
    B}_n$.

  Acting with an element of ${\cal C}_n$ changes the 7-branes in
  general. There is however a subgroup of ${\cal C}_n$ which  
  leaves the configuration invariant and thus acts as a symmetry
  of the D3 probe theory. Let us therefore define 
\item{\bf Definition:} The {\em symmetry group} ${\cal S}(w)\subset
  {\cal E}_n$ of the configuration $w$ is given by
  \begin{eqnarray}
    \label{symmdef}
    {\cal S}(w) \equiv \{(g,b)\in {\cal E}_n | \;\;
    (g,b)w=w\}\,.
  \end{eqnarray}
  ${\cal S}$ is manifestly a group, indeed a very large one because for 
   a given $g$ there typically are many choices of $b$ satisfying the    
   condition $(g,b)w =w$. 
For a given $g$
  the transformation $b$ is not unique because there are crossing
  transformations $b$ that leave $w$ invariant. These form a normal
  subgroup ${\cal H}(w)$ of ${\cal S}(w)$:
  \begin{eqnarray}
    \label{Hdef}
    {\cal H}(w) \equiv \{(\unit,b) \in {\cal S} (w)\} . 
  \end{eqnarray}
The subgroup of the IIB duality group
$\sl2z$ which
  preserves $w$ consists of all those elements $g\in\sl2z$ appearing
  in ${\cal S}(w)$. This means forgetting about the compensating transformation
$b$ and thus leads one to define: 

\item{\bf Definition:} The {\em duality group} ${\cal D}(w)$ of a
  configuration $w$ is defined as
  \begin{eqnarray}
    \label{dualdef}
    {\cal D}(w) \equiv {\cal S}(w)/{\cal H}(w)
    \cong \{g\in\sl2z | \;\;
    \exists\;b\in{\cal B}_n \;\;\mbox{such that}\;\;\;
    (g,b)w=w\}.
  \end{eqnarray}
This is clearly the subgroup of $\sl2z$ whose
  elements leave $w$ invariant after the action of a suitable
crossing transformation. 
\item{\bf Proposition:} $K(w)\in{\cal D}(w)$: the duality group
  necessarily contains the overall monodromy. In addition, $-
  \unit\in{\cal D}(w)$.
\item[]{\em Proof:} Consider the configuration $w = [A_1][A_2]\ldots
  [A_n],$ with $K_n = K(w)= A_n\ldots A_1$.  Also define $K_i =
  A_i\ldots A_1$, for $1 \leq i \leq n$.  We will explicitly construct
  an element $b$ of ${\cal B}_n$ in terms of transpositions which
  satisfies $(K_n,b)w = w$. Let us perform the $\sl2z$ transformation
  with $K_n^{-1}$:
  \begin{eqnarray}
    [A_1][A_2]\ldots [A_n]&\transform{K_n^{-1}}&
    [K_n^{-1}A_1K_n][K_n^{-1}A_2K_n]\ldots [K_n^{-1}A_nK_n] \,.
    \label{firstsl2z}
  \end{eqnarray}
  As the first step, apply the product $(P_1\ldots P_{n-1})\in\Bn$ to
  the rhs of \myref{firstsl2z}: this corresponds to moving the cut of
  the rightmost brane through the rest of them; then repeat this
  process $n-1$ times. We claim that not only the order of the branes
  is restored but their charges are transformed back to the original
  ones:
  \begin{eqnarray}
    &\transform{P_1\ldots P_{n-1}}& 
    [K_n^{-1}A_nK_n][K_n^{-1}A_nA_1A_n^{-1}K_n]\ldots 
                    [K_n^{-1}A_nA_{n\!-\!1}A_n^{-1}K_n] \nn \\
    &&=[K_{n\!-\!1}^{-1}A_nK_{n\!-\!1}]
       [K_{n\!-\!1}^{-1}A_1K_{n\!-\!1}]\ldots 
                    [K_{n\!-\!1}^{-1}A_{n\!-\!1}K_{n\!-\!1}] \nn \\
    &\transform{P_1\ldots P_{n-1}}& 
    [K_{n-1}^{-1}A_{n-1}K_{n-1}]
    [K_{n-1}^{-1}A_{n-1}A_nA_{n-1}^{-1}K_{n-1}]\ldots 
    [K_{n-1}^{-1}A_{n-1}A_{n\!-\!2}A_{n-1}^{-1}K_{n-1}] \nn \\
    &&=[K_{n-2}^{-1}A_{n-1}K_{n-2}]
    [K_{n-2}^{-1}A_nK_{n-2}]\ldots 
    [K_{n-2}^{-1}A_{n\!-\!2}K_{n-2}] \nn \\
    &&\vdots \nn\\
    &\transform{P_1\ldots P_{n-1}}&
    [A_1][A_2]\ldots [A_n]. 
  \end{eqnarray}
  Thus we proved that the overall monodromy is indeed in the duality
  group, that is   
  \begin{eqnarray}
    (K(w) ,(P_1P_2\ldots P_{n-1})^{-n})w=  (K(w)^{-1},(P_1P_2\ldots
    P_{n-1})^n)w=w.
    \label{trivial}
  \end{eqnarray}
\item[]For the transformation $g= -\unit$ we simply note that this
  transformation does not change the word describing the configuration
  since for each brane $[A_i] \to [gA_ig^{-1}] = [A_i]$. This
  transformation is clearly in the duality group and since it plainly
  leaves the 7-brane configuration invariant it does not need to be
  accompanied by brane transpositions ($b$ can be taken to be the
  identity in \myref{dualdef}).
\end{itemize}

\subsection{The action on invariant charges}

Having seen how the $\sl2z$ equivalence actions transform the 7-brane
configurations we would like to know the fate of junctions ending on
these 7-branes. A junction is characterized by its {\it invariant
  charges}; the effective number of prongs on each 7-brane of the
configuration \cite{DZ}. The action of an overall $\sl2z$
transformation is trivial: the charges of each string composing the
junction transform as a doublet but the invariant charges do not
change.

In general when one performs a crossing transformation, not only the
7-brane labels change but the invariant charges on those branes
change
as well.  This is most easily seen from the active viewpoint: instead
of relocating the cuts moving them through the 7-branes, we can move
the 7-branes. Whenever a 7-brane in motion crosses a string segment,
additional prongs on that brane might be created.

It suffices to determine how the charges change in the transposition
of two consecutive 7-branes, more complicated cases are considered by
successive transpositions. Consider therefore a junction on a pair of
7-branes $[K_{\bz_1}][K_{\bz_2}]\equiv [\bz_1][\bz_2]$ with invariant
charges $[Q_1][Q_2]$. Here we label the branes, as in \cite{finite}
with their charge vector $\bz = (p,q)$, 
in terms of which the corresponding
$\sl2z$ monodromy 
matrix is written as $K_{\bz}\equiv \unit+\bz\bz^TS$.
According to \myref{transposition} the charges of the branes
transform
as ($s_{12} \equiv \bz_1\times \bz_2= p_1q_2 - q_1 p_2$):  
\begin{eqnarray} 
  \begin{array}{lllllll}
    P&:&
    [\bz_1][\bz_2]&\mapsto& [\bz_2][K_{\bz_2}\bz_1] &=& 
    [\bz_2][\bz_1+s_{12}\bz_2] \\
    P^{-1}&:&
    [\bz_1][\bz_2]&\mapsto& [K_{\bz_1}^{-1}\bz_2][\bz_1] &=& 
    [\bz_2+s_{12}\bz_1][\bz_1]. 
  \end{array} 
  \label{transpositionz}
\end{eqnarray}
The action on the invariant charges can be determined by looking at
how many prongs are created/annihilated on each 7-brane in the
canonical presentation \cite{DZ}, but charge conservation alone gives
the answer too:
\begin{eqnarray} 
  \begin{array}{lllllll}
    P&:&
    [Q_1][Q_2]&\mapsto& [Q_2-s_{12}Q_1][Q_1] \\
    P^{-1}&:&
    [Q_1][Q_2]&\mapsto& [Q_2][Q_1-s_{12}Q_2].
  \end{array} 
  \label{transpositionQ}
\end{eqnarray}
Appending the invariant charges as superscripts to the branes, 
the complete action for 
both transpositions is:
\begin{eqnarray} 
\label{firstm}
  [{\,\bz_1}]^{\,Q_{1}}\,\, [{\,\bz_2}]^{\,Q_{2}}
  & \stackrel{P}{\longmapsto} &
  [\, \bz_{2}]^{\,Q_2 - s_{12}\,Q_1} \,\, 
  [{\bz_1 + s_{12}\, \bz_2}]^{\,Q_{1}} \\
\label{secondm}
  [{\,\bz_1}]^{\,Q_{1}}\,\, [{\,\bz_2 }]^{\,Q_{2}} 
  & \stackrel{P^{-1}}{\longmapsto} & 
  [{\,\bz_2 + \, s_{12}\, \bz_1 }]^{Q_{2}} \,\,
  [{\bz_1}]^{\, Q_1 - \, s_{12} \, Q_2}.
\end{eqnarray} 

The crossing transformation that restores the original brane
configuration after application of the monodromy $K$ (see
\myref{trivial}) has an important property: {\it it does not change
  the invariant charges of junctions with zero asymptotic charge}.
This is best seen by visualizing the motion of the 7-branes. In
\figref{noasym} we show the effect of the cyclic transformation
$P_1\ldots P_{n-1}$: it corresponds to moving the rightmost 7-brane
above the rest to the left. If a junction has no asymptotic charge,
this 7-brane does not cross any string segment along the
transformation and therefore the invariant charge of any of 7-branes
does not change. Performing this transformation $n$ times rearranges
the 7-branes in the original order and although their monodromies
change, the invariant charges of any given junction without asymptotic
charge remain the same.  
\onefigure{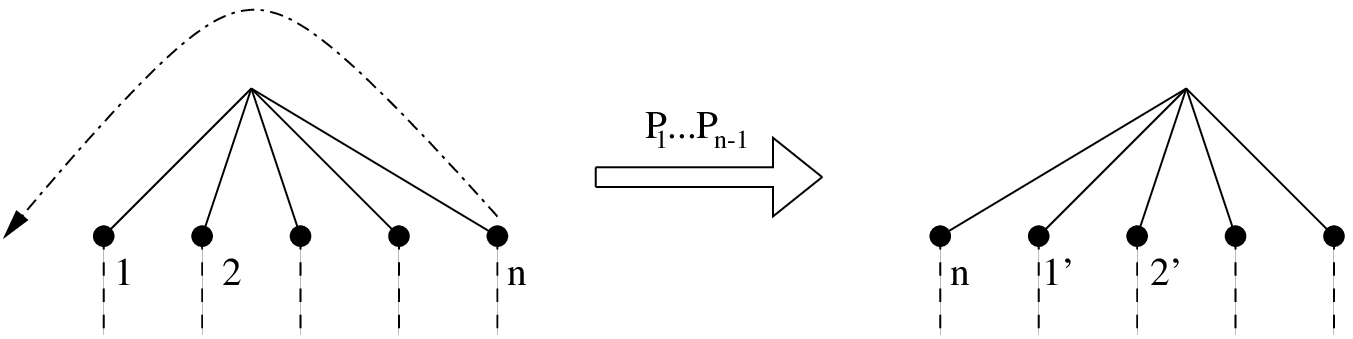}{Action of $P_1\ldots P_{n-1}$ on a 7-brane
  configuration with a string junction of no asymptotic charge.  While
  moving the rightmost brane, the $[p,q]$-charges of the branes change
  but the invariant charges of the junction do not.}

We note in passing that while the $g= -\unit \in \D(w)$ transformation
does not change the list of branes, the invariant charges of any
junction will change sign.


\section{Implementing Lie algebra Weyl transformations}
\setcounter{equation}{0}

In section 2 we introduced the group ${\cal H}(w)$ of invariant
crossing transformations. 
The elements of this group are crossing transformations 
$b$ that leave the brane configuration $w$ invariant,
namely, $bw =w$.  It is the purpose of this section to understand
some of the structure of ${\cal H}(w)$.

We know that 
invariant crossing transformations 
act on junctions by shuffling
their invariant charges, and therefore act on weight vectors
of the Lie algebra carried by the branes. We will show that
Weyl transformations of the Lie algebra can always be implemented
at the level of junctions by invariant 
crossing transformations 
and this will take most of the work in the present section.
Nevertheless, there are sometimes invariant transformations 
that do not correspond to Weyl transformations, but rather correspond
to outer automorphisms of the root lattice of the algebra.  This we
found to be a surprise.  We will give here a nontrivial example of
this phenomenon.

It is useful to define the subgroup ${\cal H}_W(w)$ called the
{\it group of invariant crossing transformations 
of Weyl type}. A transformation 
is said to belong to ${\cal H}_W(w)$ if its action on  weights
corresponding to junctions is a Weyl transformation.  There are
some special invariant transformations 
that do not change junctions at all. Such transpositions belong to
${\cal H}_W(w)$ since they imply a trivial identity Weyl
transformation.  We give an example of such transposition.

As we will see later in this paper, whenever ${\cal H}_W(w)$ coincides
with ${\cal H}(w)$, namely, if all invariant transpositions are of 
Weyl type, the characterization of duality groups is very much
simplified.  This will be the case for finite algebras, but not 
the case for affine ones.

\subsection{Weyl transformations as invariant 
crossing transformations}

In this section our main objective is to prove that for any Weyl
transformation of the Lie algebra ${\cal G}(w)$ carried by a 7-brane
configuration $w$ there is a crossing transformation which implements
this Weyl transformation at the level of junctions. This crossing
transformation leaves the brane configuration invariant and simply
acts on the invariant charges of junctions supported on the
configuration. This action is such that the associated weight vectors
undergo the desired Weyl transformation. We restrict our attention to
the finite and affine Kac-Moody algebras that can be obtained on
localizable brane configurations.

We begin by noting that Weyl transformations are generated by
elementary reflections using the simple roots of the algebras in
question. We also recall \cite{GZ,DZ,dewolfe} that each simple root
junction of the ${\bf A}^{N}, {\bf H}_{0\leq N\leq 3}, {\bf D}_{N\geq
  0}, {\bf E}_{6\leq N\leq 8}$ and $\widehat{\bf E}_{0\leq N\leq 8}$
configurations can be represented by an {\it open string} $\malpha$
($\malpha^2 = -2$) stretched between two possibly mutually non-local
7-branes.

We now claim that there is a rearrangement of the configuration,
equivalent to a crossing transformation $b$, such that the root
$\malpha$ in question becomes an open string $b(\malpha)$ stretched
between two {\it mutually local} 7-branes.  This rearrangement
corresponds to moving one of the two 7-branes on which $\malpha$ ends
along the path of $\malpha$ itself until the brane is just to the side
of other brane. This motion, for a particular example, is shown in
\figref{weyl}.  One can imagine the branch cut of the moving brane
staying vertical, and one can see that indeed this motion simply
corresponds to a sequence of transpositions of branes. After this
motion, however, many of the branes of the configuration may have
changed identity, and therefore the configuration has not been left
invariant.

\onefigure{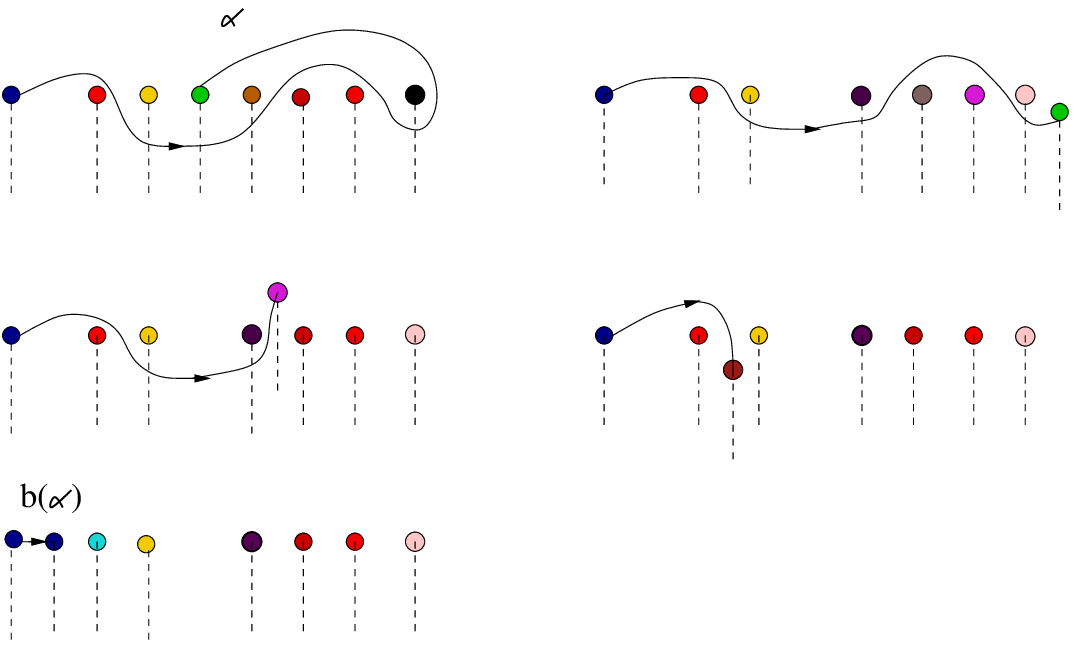}{An open string between two 7-branes can be
  transformed into an open string between two mutually local 7-branes
  by rearranging the branes.}

We now claim that the elementary Weyl reflection using the root
$\malpha$ is obtained by first doing the transpositions in $b$, then
doing the transposition $P_i$ that interchanges the two mutually local
7-branes supporting the now short open string $b(\malpha)$, and then
using the brane that sits where the first one ended to retrace
backwards the path, this is simply done by applying $b^{-1}$. Two
facts should be noted.  First, the original open string $\malpha$
changes direction, thus becoming $-\malpha$ as one would expect for a
Weyl transformation generated by $\malpha$. Second, by retracing the
path, all changes of brane labels that the first tracing of the path
caused are compensated and the brane configuration is now left
invariant. It is now left to show that this sequence of operations
$(b^{-1}P_i b)$ performs the expected Weyl reflection on arbitrary
junctions supported on the configuration. That is,
\begin{equation}
  \label{wp}
  \mbox{W}_{\alpha} = b^{-1}P_i b:\,\, {\bf J }\rightarrow {\bf J}  +
  ({\bf J} \cdot \malpha)\,\malpha \,. 
\end{equation}

We first show explicitly that this formula holds in the special case
when $\malpha$ is an open string stretching between two adjacent
mutually local branes $[{\bf z_i}]$ and $[{\bf z}_{i+1}]$.  In this
case $\malpha = {\bf z_i} -{\bf z_{i+1}}$,  
and $W_\alpha= P_i$, just
the exchange of the two branes $[{\bf z_i}]$ and $[{\bf z}_{i+1}]$.
Consider now a general junction supported on the configuration
\begin{equation}
  \label{smp}
  {\bf J} = \sum Q_k {\bf z}_k = 
  {\bf z_i} Q_i +{\bf z_{i+1}} Q_{i+1} + \cdots
  \quad \to \quad ({\bf J}\cdot \alpha ) = 
  Q_{i+1} - Q_i \,. 
\end{equation}
The exchange $P_i$ of the mutually local branes $[{\bf z_i}]$ and
$[{\bf z_{i+1}}]$ maps ${\bf J}$ to the new junction ${\bf J'}$
defined as
\begin{equation}
  \label{wpp}
  P_i: \,\, {\bf J } \to {\bf J'}  = {\bf z_i} Q_{i+1} +{\bf z_{i+1}}
  Q_{i} +
  \cdots = {\bf J} + ({\bf J} \cdot {\bf\malpha})\,{\malpha} \,,
\end{equation}
where use was made of the explicit expression for $\malpha$ and of
equation \myref{smp}.  This confirms our claim for this special case.

Let us now return to the general problem and compute the action of
$b^{-1}P_ib$ on a general junction as follows
\begin{equation}
  (\,b^{-1} P_i \,b )\, ({\bf J}) = 
  b^{-1} \, \bigl[ P_i (\, b({\bf J})) \bigr]
  = b^{-1} \bigl[\, b({\bf J}) + 
  ( b({\bf J})\cdot b (\malpha)) \, b(\malpha)\, \bigr]\,,
\end{equation}
where in the last step we used \myref{wpp} where the role of $\malpha$
is played here by the junction $b(\malpha)$ extending between the two
mutually local branes that are transposed. Since $b^{-1} b$ equals the
identity on any junction, and the intersection of two junctions is
invariant under crossing transformations we find
\begin{equation}
  (\,b^{-1} P_i \,b )\, ({\bf J})  
  = {\bf J} + ( {\bf J}\cdot \malpha) \,\malpha \,. 
\end{equation}
This completes our proof that Weyl transformations can be realized as
crossing transformations that leave invariant the brane configuration.
Our realization has been very specific, and while all such
crossing transformations belong to ${\cal H}_W(w)$, they do not
necessarily exhaust it, as we illustrate in section 3.2.

One can wonder if there is a useful notion of Weyl reflections of the
junction lattice of a configuration that makes no reference to the Lie
algebra carried by the configuration.  For any junction ${\malpha}$
such that $(\malpha\cdot\malpha)=-2$ we could define
\begin{eqnarray}
  {\cal W}_{\malpha}(\mJ)=\mJ + (\mJ\cdot\malpha){\malpha}\,.
\end{eqnarray}
This transformation preserves intersection numbers and it is therefore
an automorphism of the junction lattice. The transformations generated
by junctions $\malpha$ of zero asymptotic charge correspond to the Lie
algebraic Weyl transformations since such junctions are roots. On the
other hand a transformation ${\cal W}_{\malpha}$ generated by a
junction $\malpha$ with asymptotic charge will change the asymptotic
charge of the junction on which it acts.  The significance of such
transformations is unclear since they generically map BPS junctions to
non-BPS junctions.\footnote{Consider the $\widehat{\bf E}_{9}$
  junction $\malpha =\sum_{i=1}^{8}\mu_{i}{\omega}^{i}
  +2{\omega}^{p}-{\delta}^{(0,1)}$, with the $E_8$ weight satisfying
  $\mu^{2}=2$.  This junction has self-intersection minus two and
  non-zero asymptotic charge.  Consider now ${\cal W}_{\malpha}$
  acting on the BPS junction ${\delta}^{(0,1)}$. One readily finds
  ${\cal W}_{\malpha}(\delta^{(0,1)})=2\sum_{i=1}^{8}\mu_{i}
  \omega^{i}+ 4\omega^{p}-\delta^{(0,1)}\,.$ We know that a necessary
  condition for a junction $\mJ$ of asymptotic charge $(p,q)$ to be
  BPS is that $\mJ\cdot \mJ \geq -2+\mbox{gcd}(p,q)$ \cite{DHIZ}.
  Since $({\cal W}_{\malpha}(\delta^{(0,1)}))\cdot ({\cal W}_{\malpha}
  (\delta^{(0,1)}))=0 < -2+\mbox{gcd}(4,0) \,, $ it follows that
  ${\cal W}_{\malpha}({\delta}^{(0,1)})$ cannot be BPS.}  Thus it may
be that the only useful reflections of the junction lattice are those
generated by roots of the Lie algebra carried by the configuration.

\subsection{Further examples of invariant crossing transformations}

Above we presented a  group of invariant crossing transformations
that act on junctions via Weyl transformations of their Lie-algebraic
data.  This group, however,
does not contain all invariant transpositions ${\cal H}(w)$ (see 
(\ref{Hdef}))of the
generic 7-brane configuration. Among the additional invariant crossing
transformations there are ones that are of Weyl type as well as
others which are not. In the following we give an example for each.

{\bf ${\bf D_N}$:} For simplicity, consider first the case
of ${\bf D_1}$. This
configuration has no root junction (a junction of self-intersection minus
two and zero asymptotic charge) therefore there is no Weyl group of the
usual kind, and there are no transpositions of the type discussed in the
previous subsection. Nevertheless there is a nontrivial crossing
transformation leaving the configuration invariant and thus
belonging to ${\cal H}(w)$. This transformation leaves all
charges unchanged, as will be shown now. Indeed, 
\begin{eqnarray}
{\bf A}^{Q_{A}}{\bf B}^{Q_{B}}{\bf C}^{Q_{C}}
\transform{P_{1}} {\bf B}^{Q_{B}+Q_{A}}
{\bf X_{[0,1]}}^{Q_{A}}{\bf C}^{Q_{C}}
\transform{P_{2}} {\bf B}^{Q_{B}+Q_{A}}{\bf
C}^{Q_{C}+Q_{A}}{\bf A}^{-Q_{A}} 
\,\,\,\,\,\,\,\,\,\,\,\,\,\,\, \nonumber\\
\,\,\,\,\,\, 
\transform{P_{2}}{\bf B}^{Q_{B}+Q_{A}}{\bf A}^{Q_{C}}{\bf
X_{[0,1]}}^{Q_{C}+Q_{A}}
\transform{P_{1}}{\bf A}^{-Q_{A}+Q_{C}-Q_{B}}
{\bf X_{[2,-1]}}^{Q_{B}+Q_{A}}{\bf X_{[0,1]}}^{Q_{C}+Q_{A}}\\
\transform{P_{1}P_{2}P_{2}P_{1}}{\bf A}^{Q_{A}}
{\bf X_{[3,-1]}}^{Q_{C}}{\bf
X_{[-1,1]}}^{2Q_{C}-Q_{B}}
\transform{P_{2}}{\bf A}^{Q_{A}}{\bf B}^{Q_{B}}{\bf
C}^{Q_{C}}\,.\nn 
\,\,\,\,\,\,\,\,\,\,\,
\,\,\,\,\,\,\,\,\,\,\,\,\,\,\,
\,\,\,\,\,\,\,\,\,\,\,\,\,
\,\,\,\,\,\,\,\,\,\,\,
\end{eqnarray}
This transposition  is obtained by first 
moving the ${\bf A}$-brane around the ${\bf
BC}$ branes twice and
then moving the transformed ${\bf B}$-brane through the cut of the
${\bf C}$-brane. This brane configuration has this
particular invariance 
because $T^{2}\in {\cal D}(w)$ (section 5.3),
and therefore the $\sl2z$ transformation $T^{2k}$ induced on the ${\bf BC}$
system by the ${\bf A}$-brane can be undone by a transposition.

Now, we can turn to the case of ${\bf D_N}$, where clearly
the same transformation exists leaving the configuration and
all invariant charges unchanged. This transformation is trivially
of Weyl type in that it acts on the junctions's Lie algebraic
data as the identity. On the other hand, it is also clear that
this transformation cannot be obtained by composition of 
transformations that interchange branes connected by an open 
string.  This is the case because all such open strings in
${\bf D_N}$  join ${\bf A}$ branes and thus
the ${\bf B}$ and ${\bf C}$ branes are never interchanged.

{\bf ${\bf \widehat{\tilde{E}}_{1}}$:} This 7-brane configuration 
also does not admit any ordinary root junction.  The only
BPS junctions are multiples of the delta junction $\delta$ that
encircles the configuration.  Having no real roots we have
no transformations generated by open strings. Surprisingly, 
there is an invariant transformation that actually
changes the Lie algebraic data of junctions.  In doing so
it shows that elements of ${\cal H}(w)$ may generate
in general nontrivial automorphisms that are not of Weyl type. 
It would be of interest to exhibit those explicitly for
affine configurations having a nontrivial Weyl group, but
we focus here our attention to the case of 
${\bf \widehat{\tilde{E}}_{1}}$.  The transformation in
question is actually analogous to the one just considered above.
We first perform the following transpositions:
\begin{eqnarray}
{\bf A}^{Q_A}{\bf X}^{Q_1}_{[2,-1]}{\bf X}^{Q_2}_{[-1,2]}{\bf
X}^{Q_3}_{[1,1]}
\transform{P_1P_2P_3P_3P_2P_1} {\bf A}^{Q_A-Q_1+2Q_2+Q_3}{\bf
X}^{Q_1+Q_A}_{[3,-1]}
{\bf X}^{Q_2+Q_A}_{[-3,2]}{\bf X}^{Q_3-Q_A}_{[0,1]}\,.
\end{eqnarray}
This transposition consists of moving the ${\bf A}$-brane around
the other 7-branes once in the counter clockwise direction. The 
linear transformation $g$ induced on the charges is given by
\begin{eqnarray}
\left( \begin{array}{cc} Q_A\\ Q_1 \\ Q_2 \\ Q_3 \end{array} \right )
\mapsto
 \left( \begin{array}{cccc} ~~1 & -1 & 2 & 1\\ ~~1 &~~1 & 0 
& 0\\ ~~1 &~~0 & 1 &0\\ -1 &~~0 & 0 &1 \end{array} \right )
\left(\begin{array}{c}
Q_A \\ Q_1 \\ Q_2 \\ Q_3 \end{array} \right) \,.
\label{gg}
\end{eqnarray}
The effect on the charges of moving the ${\bf A}$-brane around the other 
7-branes three times is given by the linear transformation $g^{3}$. We can
restore the original configuration by transposition $P_2P_3$,
\begin{eqnarray}
{\bf A}^{Q_A}{\bf X}^{Q_1}_{[2,-1]}{\bf X}^{Q_2}_{[-1,2]}{\bf
X}^{Q_3}_{[1,1]}
\transform{(P_{1}P_{2}P_3P_3P_2P_1)^{3}} {\bf A}^{Q_A'''}{\bf
X}^{Q_1'''}_{[5,-1]}{\bf X}^{Q_2'''}_{[-7,2]}{\bf
X}^{Q_3'''}_{[-2,1]}\transform{P_2P_3}{\bf A}^{\widehat{Q}_A}{\bf
X}^{\widehat{Q}_1}_{[2,-1]}{\bf X}^{\widehat{Q}_2}_{[-1,2]}{\bf
X}^{\widehat{Q}_3}_{[1,1]}\,.\nn
\end{eqnarray}
The transformed charges $\widehat{Q}$ are
\begin{eqnarray}
\widehat{Q}_A=Q_A'''\,,\, \widehat{Q}_{1}=Q_3'''+3Q_2'''-3Q_1'''\,,\,
\widehat{Q}_2=Q_1'''\,,\,\widehat{Q}_3=-Q_2'''\,,
\end{eqnarray}
where $\vec{Q}'''$ are obtained from $\vec{Q}$ by the linear
transformation $g^3$. Using (\ref{gg}) we get
\begin{eqnarray}
\widehat{Q}_A&=&Q_A-3Q_1+6Q_2+3Q_3\,,\nn\\
\widehat{Q}_{1}&=&3Q_A+3Q_2+2Q_3\,,\nn\\
\widehat{Q}_2&=&3Q_A-2Q_1+6Q_2+3Q_3\,,\\
\widehat{Q}_3&=&-3Q_A+3Q_1-7Q_2-3Q_3\,.\nn
\end{eqnarray}
Thus we see that $(\unit,P_2P_3(P_1P_2P_3P_3P_2P_1)^3)\in {\cal
H}(w)$ has a nontrivial action on the charges. Since there are no real
root junctions with support on this configuration this
element of ${\cal H}(w)$ is not of Weyl type. The motion of the ${\bf
A}$-brane
around the other three branes has the effect equivalent to the action
of a global $\sl2z$ transformation by $T^{-3}$.  We will discuss this
action in more detail 
in section 6.


\section{Duality groups and Dynkin graph automorphisms}
\setcounter{equation}{0}
The results of the previous section allow us to find and characterize
the duality group $\D(w)$ of a brane configuration $w$ with monodromy
$K(w)$. Since crossing transformations cannot change the monodromy
$K(w)$ of the configuration, any element of the duality group must
leave the monodromy invariant and therefore is contained in Stab$(K)$,
the stabilizer of $K$ in $\sl2z$ ($g\in \hbox{Stab}(K) \Leftrightarrow
gKg^{-1} = K$)
\begin{equation}
  \label{fcond}
  \D (w ) \subseteq \hbox{Stab} \, (K(w) )\,.
\end{equation} 
The duality group $\D(w)$ of the configuration will be the subgroup of
Stab$(K)$ that leaves the configuration invariant in the sense
discussed in the previous section (see \myref{dualdef}).  In any
concrete case it is relatively straightforward to determine the group
Stab$(K)$.  Then one must select $\D$ by finding the subgroup of
Stab$(K)$ for which there exist crossing transformations that restore
the configuration.

We have seen that $K \in \D$ (Proposition, sect.~2.1) and therefore
$\{K \}$, the group generated by $K$, is a subgroup of the duality
group $\D$.  Since $\{K\}$ is a normal subgroup of Stab$(K)$, it is
also a normal subgroup of $\D$.  We are thus led to define the
quotient group
\begin{equation}
  \label{quot}
  \overline \D \equiv \D/\{K\}\,,
\end{equation} 
referred to as the {\it reduced} duality group, that will play an
important role in the computations.  Another general fact discussed
before is that the transformation $-\unit \in \sl2z$, clearly
contained in Stab$(K)$ is also an element in $\D$.

Consider now the action of an element $(g,b)\in {\cal S}(w)$ of the
symmetry group of the brane configuration.  The
 set of transpositions
$b$ that restore the original configuration via $(g,b)w =w$
will shuffle the invariant charges
of junctions and therefore this symmetry maps junctions to
(typically) different junctions.  One nevertheless gets an
automorphism of the junction lattice $\Lambda_{\bf J}$; namely for any
two junctions $\bf J_1$ and $\bf J_2$ mapping to ${\bf J}_1'$ and
${\bf J}_2'$ one has ${\bf J}_1\cdot {\bf J}_2={\bf J}_1'\cdot{\bf
  J}_2' $. We therefore have a map
\begin{equation}
\label{needl}
{\cal S} \to \hbox{Aut} (\Lambda_{\bf J})\,,
\end{equation}
from the symmetry group of the brane configuration
 to the group of automorphisms of the
junction lattice.

Let us focus for the moment on junctions representing roots of the Lie
algebra associated to the brane configuration.  In fact, more
generally, consider junctions associated to elements in the {\it root
  lattice} $Q$ of the Lie algebra. Such junctions, as discussed in
length in earlier papers \cite{DHIZ} have zero asymptotic charges.  It
follows that 
symmetry transformations in ${\cal S}$ 
will map these junctions among
themselves.  In addition, for such junctions ${\bf J}_1 \cdot {\bf
  J}_2 = - \lambda_1 \cdot \lambda_2$, where $\lambda_1, \lambda_2 \in
Q$ are the associated elements in the root lattice. Since duality
elements map to automorphisms of the junction lattice, by restricting
to junctions associated to $Q$ duality elements map to automorphisms
of the root lattice $Q$. We therefore have:
\begin{equation}
  \label{needq}
  {\cal S} \to \hbox{Aut} (Q)\,.
\end{equation}

On the other hand  there is no
canonical map $\D\to$ Aut$(\Lambda_{\bf J})$, nor there is a canonical
map $\D\to$ Aut$(Q)$ from the duality
group of the configuration.  This is so because for duality elements, 
the compensating
crossing transformation in ${\cal H}(w)$ used to restore the brane configuration
is not uniquely defined. 
 As discussed in section 3, a brane
configuration typically admits crossing transformations $\widehat b$
that leave it invariant. In fact, any Weyl reflection of the root
lattice is generated on junctions by a crossing transformation
$\widehat b$ that leaves the configuration invariant. 
In the language of \myref{dualdef} the
ambiguous action of a duality transformation $g$ on junctions arises
because the computation of such action requires the choice of some $b$
such that $(g,b)w=w$. But $b$ is ambiguous, if $b$ satisfies this
equation, $b\widehat b$ does as well.

If a configuration $w$ has the property that ${\cal H}(w)={\cal
  H}_{W}(w)$, namely,  every invariant transposition is of Weyl type, then
the map  
\begin{eqnarray}
\label{fidu}
{{\cal D}}\to \mbox{Aut}(Q)/W\,.
\end{eqnarray}
is well defined.  This is the case because all invariant transpositions
map to the Weyl group. Since the duality
elements
$\{K\}$ always map to Weyl group we also have the well defined map
\begin{eqnarray}
\label{fiduu}
{\overline{\cal D}}\to \mbox{Aut}(Q)/W\,.
\end{eqnarray}
In the above homomorphisms the  quotient
group to the right  is well defined since
$W$ is a normal subgroup of $\hbox{Aut}(Q)$ (for this and other facts
quoted below see, \cite{kac}). This quotient, for algebras of finite
type, is simply the (graph) automorphism group $\Gamma$ of the Dynkin
diagram (and the $\pm$ above is not necessary).  For infinite
Kac-Moody algebras the above quotient includes, in addition to the
graph automorphism $\Gamma$, the generator $(-1)$ which changes the
sign of every vector in the root lattice, and this is never a Weyl
transformation. For finite algebras the transformation $(-1)$ of $Q$
is many times a Weyl transformation. From the list of finite algebras
we consider, $(-1)$ is not a Weyl element for the $A_N$ series and for
$E_6$.  In such cases, $(-1)$ is equivalent, up to Weyl
transformations, to a graph automorphism.  Since the element $-\unit
\in \D \subseteq \sl2z$ precisely acts as $(-1)$ on $Q$ this shows
that this transformation is nontrivial ({\it i.e.}  not Weyl) for the
$A_N$ series, for $E_6$ and for affine algebras.

In the next section we will show that ${\cal H}(w)={\cal
H}_{W}(w)$ for 7-brane configurations 
realizing finite algebras. This condition, however, 
may not satisfied for
configurations realizing affine algebras, 
as we illustrated in section 3.2. 
Therefore in these cases
(\ref{fidu}) and (\ref{fiduu}) are
 not well defined since  there is no
unique choice of invariant transposition 
 and different choices can induce non-Weyl action on the
roots.

Our strategy in this case would be to define the maps 
${{\cal
D}}\to \pm \Gamma$ and $\overline{{\cal
D}}\to \pm \Gamma$ with respect to a fixed set of invariant
transpositions that can undo the effect of an $\sl2z$ transformation.
The map is thus dependent on the choice of transpositions.
An invariant characterization will require better understanding
of the structure of ${\cal H}(w)$.

When defined, our interest is in the homomorphisms $\D \to \pm \Gamma$ and 
 $\overline{{\cal
D}}\to \pm \Gamma$, the latter 
capturing the interplay of duality transformations with Lie algebraic
data. Dualities in $\D$ that map to nontrivial elements of $\Gamma$
relate junctions appearing in different representations or junctions
appearing as vectors in $Q$ that are not related by Weyl
transformations.  For the finite algebras, 
we shall find cases when the map $\phi
:\overline{{\cal D}}\to \pm \Gamma$ is an
isomorphism (for $ E_6$, for example), and cases when it is onto but
not one to one ($D_4$, for example).

We can readily find the implications for weight vectors in the case of
configurations leading to finite algebras.  In this situation ${\bf
  J}_1 \cdot {\bf J}_2 = - \lambda_1 \cdot \lambda_2 + f (p_1, q_1;
p_2, q_2)$, where $f$ is a quadratic form determined solely by the
monodromy $K$ \cite{DHIZ}.  Since duality transformations preserve $K$
the automorphisms of the junction lattice arising from dualities give
automorphisms of the weight lattice $\Lambda$. On the other hand for
finite algebras Aut$(Q) = $ Aut $(\Lambda)$ since $\Lambda=Q^*$. It
thus follows that the homomorphism $\D \rightarrow \pm\Gamma$ carries
information on how representations in different conjugacy classes are
mapped into each other by duality transformations. We leave the
question of duality action on junctions corresponding general affine
weight vectors open.


\section{Duality groups for finite-type configurations}
\setcounter{equation}{0}

In this section we calculate duality groups $\D$ and give the
homomorphisms to the corresponding Dynkin-graph automorphism groups
$\Gamma$.  These homomorphisms are characterized by $\phi: \overline
\D \to \Gamma$, as discussed before. This section focuses on brane
configurations of elliptic and parabolic monodromies realizing finite
Lie algebras.
The case of finite algebras is relatively simple to analyze. The
7-brane configurations realizing finite algebras have the property
that ${\cal H}(w)={\cal H}_{W}(w)$ i.e, all invariant transpositions are
of Weyl type. To prove this consider the action of a invariant
transposition on the weight vector of a junction with support on the
7-brane configuration. If the transposition is not of Weyl type it
will induce an outer automorphism on the weight vector. It was shown
in \cite{DZ} that for 7-brane configurations realizing finite algebras
the conjugacy class of a weight vector corresponding to a junction is
determined by the asymptotic charge of the junction. Since a
transposition cannot change the asymptotic charge of a junction, it
cannot change the conjugacy class of the corresponding weight
vector. Therefore the action of the transposition on the weight vector
cannot be an outer automorphism and hence must be a Weyl
transformation. This simplification implies that the homomorphism
$\overline{{\cal D}} \mapsto \Gamma$ is well defined for these
cases. In table 1 we list the relation between the conjugacy classes
and the asymptotic charge for various 7-brane configurations realizing
finite algebras. In the case of $E_{8}$ since there are no outer
automorphisms therefore every transpositions is trivially of Weyl
type. This is consistent with the fact that there is a single conjugacy
class for $E_{8}$.
\begin{table}
\begin{eqnarray*} 
\begin{array}{||c|l|c|c||} \hline
  \rule{0mm}{5mm} {\cal G} & w
  & {\cal C} & {\mbox{ constraint}}
  \\ \hline 
  \rule{0mm}{5mm} A_{N} & {\bf A^{N+1}}, 
{\bf A^{N+1}C}& \bbbz_{N} & p ~(\mbox{mod}
~N)\\ \hline
 \rule{0mm}{5mm} D_{N} & {\bf A^{N}BC}& 
\bbbz_{2}\times \bbbz_{2}, N=\mbox{even}& p-q
~(\mbox{mod}~2), q~(\mbox{mod}~2)\\ \hline
  \rule{0mm}{5mm} & & \bbbz_{4}, 
N=\mbox{odd}& 2p-q ~(\mbox{mod}~4)\\ \hline
 \rule{0mm}{5mm} E_6& {\bf A^{5}BCC}& 
\bbbz_{3}& p ~(\mbox{mod}~3)\\ \hline
\rule{0mm}{5mm} E_7& {\bf A^{6}BCC}& 
\bbbz_{2}& p +q~(\mbox{mod}~2)\\ \hline
\end{array}
\end{eqnarray*} 
\caption{Conjugacy classes of finite algebras,
their group structure ${\cal C}$,  
and the asymptotic charges
of junctions representing such conjugacy classes.}
\end{table}

Throughout this and the next section we denote by $\{ \cdots \}$ a
group generated by the elements indicated by dots. In addition $\{
\cdots | \cdots\}$ will denote the group generated by the elements to
the left of the vertical bar, modulo the relations to the right of the
bar.

\subsection{\bf ${\bf A_{N}}$ configuration: ${\bf A^{N+1}}$}
This configuration is built from $(N+1)$ $[1,0]$ branes. The monodromy
is $K=T^{-N-1}$ and $\hbox{Stab} (K) = \{-\unit,T\}$. Since $T$
preserves the charges of the 7-branes, it belongs to the duality group
just like $-\unit \in \D$.  Therefore $\D({\bf A_N}) =\mbox{Stab}(K)=
\{-\unit,T\} = \bbbz_{2}\times \bbbz_{}$, and $\overline \D({\bf
  A_N})= \{-\unit, T~|~T^{N+1}=\unit\}=\bbbz_{2}\times\bbbz_{N+1}$.

On the other hand $\Gamma (A_N)=\bbbz_{2}$ for $N\geq 2$ and is
generated by the transformation $ {\cal O} :\,\,
(a_{1},a_{2},\cdots,a_{n})\mapsto (a_{n},a_{n-1},\cdots,a_{1})$ of the
Dynkin labels.  Since $T$ does not affect the invariant charges it
leaves all Dynkin labels unchanged and therefore $\phi(T) = \unit
\in\Gamma (A_N)$.  The transformation $-\unit$, however, changes the
sign of all invariant charges and therefore of all the Dynkin labels
\begin{equation}
  -\unit :\,\, (a_{1},a_{2},\cdots, a_{n})\mapsto
(-a_{1},-a_{2},\cdots
  ,-a_{n})\,.
\label{minusone}
\end{equation}
The Weyl transformation $W$ which corresponds to rotating the 7-brane
configuration by half a full turn can be seen to map
\begin{equation}
  W:\,\, (a_{1},a_{2},\cdots ,a_{n})\mapsto 
  (-a_{n},-a_{n-1},\cdots, -a_{1})\,.
  \label{weyl-An}
\end{equation}
We now recognize that the action of $-\unit\in \sl2z$ on the Dynkin
labels is given by the composition of ${\cal O}$ and $W$. Therefore
the homomorphism $\phi$ from $\D$ to $\Gamma(A_N)$ is defined as
\begin{eqnarray} 
  \phi(T) &=& +1\in
  \Gamma (A_N) \nn\\
  \phi(-\unit)&=& {\cal  O} \in \Gamma(A_N)\,. 
\end{eqnarray} 

In case of ${\bf A_1}$ the computation of the duality group is
identical and therefore $\overline \D({\bf A_1})=
\bbbz_{2}\times\bbbz_{2}$. On the other hand here $\Gamma (A_1) =1$
and the $-\unit$ transformation is simply a Weyl transformation, thus
the homomorphism $\phi$ is trivial.

\begin{table}
\begin{eqnarray*} 
\begin{array}{|c|l|c|c|c|c|c|} \hline
  \rule{0mm}{5mm} w & K(w)
  & {\cal D}(w) &\overline{{\cal D}}(w) & {\cal G} & 
  \Gamma({\cal G})  
  \\ \hline 
  \rule{0mm}{5mm} {\bf A_{1}} & T^{-2} & \{-\unit, T\} &
  \bbbz_{2}\times\bbbz_{2} & A_{1} & 1
  \\ \hline
  \rule{0mm}{5mm} {\bf A_{N\geq 2}} & T^{-N-1} & \{-\unit, T\}
  & \bbbz_{2}\times\bbbz_{N+1} & A_{N} & \bbbz_{2}\\ \hline
\end{array}
\end{eqnarray*} 
\caption{Duality groups and graph automorphisms for ${\bf A_N}$
  configurations.}   
\end{table}

\subsection{${\bf H_N}$ configurations: ${\bf A^{N+1}C}$} 

Since ${\bf H_3} = {\bf D_3}$, and ${\bf H_{N\geq 4}}$ have hyperbolic
monodromies, we need only focus on the configurations ${\bf H_0, H_1}$
and ${\bf H_2}$.

\begin{itemize}
\item ${\bf H_0:}$ The monodromy $K({\bf H_0})\sim (ST)^{-1}$ and
  $\mbox{Stab}(K)=\{(ST)^{-1}\}$ ($-\unit\in \mbox{Stab}(K)$ since
  $(ST)^3= -\unit$). Since $K\in {\cal D}$, ${\cal D}({\bf H_0})=
  \mbox{Stab}(K)=\bbbz_{6}$, and $\overline{{\cal D}}({\bf H_0})=1$.
  This configuration supports no junctions without asymptotic charges
  so there is no ${\cal G}$ associated to it.
  
\item ${\bf H_1:}$ Here $K({\bf H_1})\sim S^{-1}$ and
  $\mbox{Stab}(K)=\{ S\}$. Since $K\in \D$, ${\cal D}({\bf
    H_1})=\mbox{Stab}(K)=\bbbz_{4}$ and $\overline{{\cal D}}({\bf
    H_1})=1$.  Since $\Gamma(A_1)=1$, $\phi$ is the trivial
  isomorphism.
\item ${\bf H_2:}$ In this case $K({\bf H_2})\sim -ST$ and
  $\mbox{Stab}(K)=\{ST\}$.  Since $(ST)^{3}=-\unit$, ${\cal D}({\bf
    H_2})=\mbox{Stab}(K)=\bbbz_{6}$ and $\overline{{\cal D}}({\bf
    H_2})=\bbbz_{2}$.  Here $\Gamma(A_2)=\bbbz_{2}$, and it follows
  from (\ref{minusone}) and (\ref{weyl-An}) that the homomorphism
  $\phi$ from $\overline{{\cal D}}({\bf H_2})$ to $\Gamma(A_{2})$ is
  given by
\begin{eqnarray}
  \phi(\unit)&=&+1 \in \Gamma(A_{2})\,,\nn \\  
  \phi(-\unit)&=&{\cal O} \in \Gamma(A_{2})\,, 
\end{eqnarray}
where ${\cal O}(a_{1},a_{2}) =(a_{2},a_{1})$.
\end{itemize}
\begin{table}
  \begin{eqnarray*} 
  \begin{array}{|c|l|c|c|c|c|c|} \hline
  \rule{0mm}{5mm} w & K(w) & {\cal D}(w) 
  &\overline{{\cal D}}(w) & {\cal G} & \Gamma({\cal G})
  \\ \hline 
  \rule{0mm}{5mm} {\bf H_0} &  ~~(ST)^{-1} & \{ST\}= \bbbz_6 & 1& -- & --   
  \\ \hline
  \rule{0mm}{5mm} {\bf H_1} &  ~~~~S^{-1} & \{S\} = \bbbz_4 & 1 & A_{1} & 1 
  \\ \hline
  \rule{0mm}{5mm} {\bf H_2} &  -(ST) &  \{ST\} =\bbbz_{6} & 
  \bbbz_{2} &    A_{2}        &      \bbbz_{2}   
  \\ \hline
  \end{array}
  \end{eqnarray*} 
\caption{Duality groups and graph automorphisms for ${\bf H_N}$
  configurations.}
\end{table}

\subsection{${\bf D_N}$ configurations: ${\bf A^{N}BC}$}   

We begin with some general remarks applicable whenever $N\not= 0$.  We
will show that $T\in \D$.  Indeed a $T$ transformation can be undone
by taking the rightmost ${\bf A}$ brane on a round trip encircling
${\bf BC}$ branes:
\begin{equation}
\label{trdn}
 {\bf A^{N-1} \, A\, BC} \transform{T} {\bf A^{N-1} \, 
   A\, X_{[0,-1]} X_{ [2, 1]} }
    \,\,\transform{P_NP_{N+1}P_{N+1}P_N} \,\,{\bf A^{N-1}\, A\, BC}\, . 
\end{equation}
In doing this operation the invariant charges on the branes, denoted
as $Q_1, \cdots Q_{N-1}$, for the inert ${\bf A}$ branes, $Q_N$ for
the rightmost ${\bf A}$ brane, and $Q_B, Q_C$, transform as
\begin{eqnarray}
  \label{gentry}
  Q_{i}&\transform{T}& Q_{i}\,, \;\;\;\;\;i=1\ldots N-1\nn\\ 
  Q_N&\transform{T}& -Q_N-Q_{B}+Q_{C} \,,\nn\\ 
  Q_{B}&\transform{T}&Q_{B}+Q_N \,, \\
  Q_{C}&\transform{T}&Q_{C}+Q_N\nn\,. 
\end{eqnarray}
On the other hand, we know form eq.(6.27) of \cite{DZ} how Dynkin
labels are given in terms of invariant charges:
\begin{eqnarray}
  \label{ggntry}
  a_{i}&=& Q_{i}- Q_{i+1} \,, \;\;\;\;\;i=1\ldots N-2\nn\\ 
  a_{N-1} &=& Q_{N-1}-Q_N \,,\\ 
  a_N&= & Q_{N-1} + Q_N + Q_{B}- Q_C  \nn\,. 
\end{eqnarray}
One immediately deduces from the last two equations the action of the
duality $T$ on the Dynkin labels
\begin{equation}
  \label{daut}
  T: \, ( a_1 , \cdots, a_{N-1}, a_N) \mapsto  
  ( a_1 , \cdots, a_{N}, a_{N-1})\,.   
\end{equation}
This exchange of the last two Dynkin labels is the familiar
$\bbbz_{2}$ automorphism of the $D_N$ Dynkin graph.

\begin{itemize}
\item ${\bf D_0:}$ The monodromy $K({\bf BC})=-T^{4}$ and
  $\mbox{Stab}(K)=\{-\unit,T\}$. A junction of asymptotic charge
  $(p,q)$ on this configuration satisfies the condition $p+q\equiv
  0~(\mbox{mod}~2)$. After a transformation by $T^{k}$ if the 
branes can be brought back to the original ones by branch cut moves then the
  transformed asymptotic charge $(p',q')=(p+kq,q)$ must also satisfy
  the same condition. This implies that $k\equiv 0~(\mbox{mod}~2)$. 
   Indeed, a probe D3-brane in this background realizes
  ${\cal N}=2$ pure SW-theory whose BPS spectrum is not invariant
  under $T$ transformation \cite{DHIZ}. One can verify, however, that
  this configuration is invariant under transformation by $T^{2}$:
  \begin{equation} {\bf BC}\,\,\, \transform{T^{2}} \,\,\,{\bf
  CX_{[3,1]}}\,\,\,\transform{P_{1}^{-1}}\,\,\, {\bf BC}\,.
  \end{equation} It then follows that ${\cal D}({\bf D_{0}})=\{-\unit,
  T^{2}\}$ and $\overline{{\cal D}}({\bf
  D_{0}})=\{-T^{2}~|~T^{4}=-\unit\}=\bbbz_{4}$.  Since this
  configuration supports no junctions without asymptotic charges there
  is no ${\cal G}$ associated to it.
  
\item ${\bf D_1:}$ Here $K({\bf D_1})=-T^{3}$ and
  $\mbox{Stab}(K)=\{-\unit,T\}$. Since we have an ${\bf A}$ brane
  $T\in \D$.  Thus ${\cal D}({\bf D_1})=\{-\unit,T\}$ and
  $\overline{{\cal D}}({\bf D_{1}})=\{T~|~T^{3}=-\unit\}=\bbbz_{6}$.
  This configuration does not support any root, therefore there is no
  Dynkin diagram. Nevertheless, as is well known, it carries a $u(1)$
  algebra, whose associated junction is the non-BPS junction
  $\overline\mJ = 2{\bf a} - {\bf b} - {\bf c}$ \cite{SZ}.  For an
  arbitrary junction $\mJ$ the corresponding $u(1)$ charge $Q^*$ is
  proportional to $\mJ \cdot \overline{\mJ} \sim 2Q_1 + Q_B - Q_C$.
  Both $-\unit$ and $T$ are checked to take $Q^* \to -Q^*$.
  
\item ${\bf D_2:}$ This configuration has $K({\bf D_2})=-T^{2}$,
  $\mbox{Stab}(K)=\{-\unit,T\} = {\cal D}({\bf D_2})$.  Therefore
  $\overline{{\cal D}}({\bf D_2})= \{T~|~T^{2}=-\unit\} = \bbbz_{4}$.
  The configuration supports two roots representing the $A_1\oplus
  A_1$ algebra, it corresponds to two disconnected Dynkin nodes, with
  Dynkin labels $a_1$ and $a_2$ correctly given by \myref{ggntry}.
  The action of $T$ as given in \myref{daut} simply exchanges the two
  Dynkin labels.  This is the non-trivial element of $\Gamma(A_1\oplus
  A_1)=\bbbz_{2}$. Therefore $\phi: \overline{{\cal D}}({\bf D_2}) =
  \bbbz_{4}\mapsto \bbbz_{2}$ via $\phi(T)=-1\,.$
 
\item ${\bf D_3:}$ Here $K({\bf D_3})=-T$,   
  $\mbox{Stab}(K)=\{-\unit,T\} = {\cal D}({\bf D_3})$, and
  $\overline{{\cal D}}({\bf D_3})=\{T~|~T=-\unit\}=\bbbz_{2}$. Also
  $T: (a_{1},a_{2},a_{3})= (a_{1},a_{3},a_{2})\,,$ is the non-trivial
  element of $\Gamma(A_3)=\bbbz_{2}$ (the labeling of nodes follows
  the $D_N$ conventions; node number one is in the middle).  Thus the
  homomorphism $\phi:\overline{{\cal D}}({\bf D_3}) \mapsto
  \Gamma(A_3)=\bbbz_{2}$ is the isomorphism $\phi(T)=-1\,.$
  
\item ${\bf D_{4}}:$ In this case the monodromy is $-\unit$ and
  therefore $\mbox{Stab}(K)=\sl2z$. Since invariance under $T$ has
  been already established, we show that ${\cal D}({\bf
    D_4})=\mbox{Stab}(K)=\sl2z$ by demonstrating the invariance of the
  configuration under $S$. Indeed,
  \begin{eqnarray}
    \label{stran}
    {\bf A^{4}BC} &\transform{S}& {\bf (X_{[0,1]})^{4}CB}
    \transform{(P_5P_4P_3P_2P_1P_1P_2P_3P_4)}
    {\bf A^{4}BC}\,. 
  \end{eqnarray}
  It is possible to anticipate the action of $S$ on the $D_4$ Dynkin
  labels.  Recall from \cite{DZ} that the various conjugacy classes of
  $so(8)$ are correlated with asymptotic $(p,q)$ charges mod 2. In
  particular in eq.~(6.26) of \cite{DZ} we see that ${\bf 8_v}$ and
  ${\bf 8_s}$ representations arise from $(1,0)$ and $(0,1)$ charges
  respectively (mod 2), while ${\bf 8_c}$ arises from $(1,1)$. We see
  that mod 2, the action of $S$ on those asymptotic charges exchanges
  the ones corresponding to ${\bf 8_v}$ and ${\bf 8_s}$ while it
  leaves invariant that corresponding to ${\bf 8_c}$. In our
  conventions, ${\bf 8_v}$ and ${\bf 8_s}$ are associated to the first
  and third nodes of the Dynkin diagram, and therefore we expect $S$
  to act as
  the graph automorphism $a_1 \leftrightarrow a_3$.  
 
  The transformations in \myref{stran} imply that under $S$, the
  invariant charges transform as:
  \begin{eqnarray}
    Q_{i}&\transform{S}&Q_{B}-Q_{i} +
    \sum Q_k\;\;\;\; i=1\ldots 4\;\;\; \nn\\ 
    Q_{B}&\transform{S}&-Q_{C}-2Q_{B}-2\sum Q_k \\
    Q_{C}&\transform{S}&-Q_{B}-\sum Q_k\nn\,. 
  \end{eqnarray}  
 
  The resulting action on the Dynkin labels \myref{ggntry} is given
  by
  \begin{eqnarray}
    S: (a_{1},a_{2},a_{3},a_{4}) &\mapsto
    &(-a_{1}\, ,-a_{2}\, ,-a_{3}\, ,a_{1}+2a_{2}+a_{3}+a_{4}) \,.
  \end{eqnarray}
  A little calculation shows that $S$ is a composition of Weyl
  reflections and the expected graph automorphism:
  \begin{equation}
    S = {\cal O} \, 
    W_{\alpha_{1}+\alpha_{2}+\alpha_{3}}W_{\alpha_{2}}\,,\qquad
    {\cal O}  
    (a_{1},a_{2},a_{3},a_{4}) \equiv (a_{3}, a_{2},a_{1},a_{4}) \,. 
  \end{equation}
\item ${\bf D_{N\geq 5}:}$ Here $K({\bf D_N})=-T^{4-N}$ and
  $\mbox{Stab}(K) = \{ -\unit , T \}= \D ({\bf D_N})$.  Thus
  $\overline \D({\bf D_N})=\{T~|~T^{N-4} = -\unit\}=\bbbz_{2(N-4)}$.
  Under $T$, using \myref{daut}, we have $\phi : \overline \D ({\bf
    D_N}) \mapsto \Gamma(D_{N})=\bbbz_{2}$ is fixed by $\phi(T)=-1$.
  Since we have a homomorphism, $\phi(-\unit) = \phi(T^{N-4}) =
  [\phi(T)]^{N-4} = (-1)^{N-4} = (-1)^N$. Thus for ${\bf D}_N$ with
  $N$ even, the transformation $-\unit$ maps to a Weyl transformation,
  while for $N$ odd, the transformation $-\unit$ is equivalent to the
  nontrivial graph automorphism up to a Weyl transformation.  This is
  as expected; a change of sign of all Dynkin labels in the $D_N$
  algebras is a Weyl transformation only for $N$ even
  (Ref.~\cite{humphreys}, sect.~13).

\end{itemize}
\begin{table}
\begin{eqnarray*} 
\begin{array}{|c|l|c|c|c|c|c|} \hline
  \rule{0mm}{5mm} w & K(w)  & {\cal D}(w) 
  &\overline{{\cal D}}(w) & {\cal G} & 
  \Gamma({\cal G})
  \\ \hline 
  \rule{0mm}{5mm} {\bf D_0}& -T^4 & \{-\unit,  T^2\} 
  & \bbbz_4 & -- & -- 
  \\ \hline
  \rule{0mm}{5mm} {\bf D_1}& -T^3 & \{-\unit, T\}& \bbbz_6 & u(1) 
  & --
  \\ \hline
  \rule{0mm}{5mm} {\bf D_2}& -T^2 & \{-\unit,  T\}& \bbbz_4 
  & A_1\oplus A_1& \bbbz_2   
  \\ \hline
  \rule{0mm}{5mm} {\bf D_3}& -T   & \{-\unit,  T\}& 
  \bbbz_2 & A_3& \bbbz_2  
  \\ \hline
  \rule{0mm}{5mm} {\bf D_4}& -\unit& \sl2z & P\sl2z &D_4 & S_3
  \\ \hline
  \rule{0mm}{5mm} {\bf D_{4+N\geq 5}}    & -T^{-N} & \{-\unit, T\} &
  \bbbz_{2N} & D_N & \bbbz_{2}
  \\ \hline
\end{array}
\end{eqnarray*} 
\caption{Duality groups and graph automorphisms for ${\bf D_N}$
  configurations. $S_{3}$ is the permutation group of three objects. } 
\end{table}


\subsection{$\bf E_{N}$ configuration: $\bf A^{N-1}BCC$}

\begin{itemize}
\item ${\bf E_{6}}$: Here $K({\bf {E_{6}}})\sim -(ST)^{-1}$ and
  $\mbox{Stab}(K) =\{-\unit, K\} = \bbbz_6= \D({\bf E_6})$.  Therefore
  $\overline \D({\bf E_{6}})= \{ -\unit\}= \bbbz_2$. In addition,
  $\Gamma(E_{6})=\bbbz_{2}$, and its non-trivial element, up to a Weyl
  transformation, changes the sign of all the Dynkin labels (maps
  representations to their conjugates).  It follows that the
  homomorphism $ \phi: \overline \D({\bf E_{6}} )\mapsto \bbbz_{2}$ is
  fixed by $\phi(-\unit)=-1$. This is an isomorphism.
\item ${\bf E_{7}}$: Here $K({\bf E_{7}})\sim S$ and
  $\mbox{Stab}(K)=\{S\}=\bbbz_4 = \D({\bf E_{7}})$. It follows that
  $\overline \D(E_{7})=\{\unit\}$.  Since $\Gamma(E_{7})$ is also
  trivial the homomorphism $\phi$ is trivial.  Dualities will preserve
  Weyl orbits, and therefore representations.  While the duality
  $-\unit = S^2$ changes the sign of all Dynkin labels, this is simply
  a Weyl transformation of $E_7$.
\item ${\bf E_{8}}$: Here $K({\bf E_8}) \sim (ST)^{-1}$ and
  $\mbox{Stab}(K)=\{(ST)^{-1}\}=\bbbz_6 = \D({\bf E_{8}})$.  Just as
  in the case of $E_7$ we have $\overline\D({\bf E_{8}})= \{\unit\}$,
  $\Gamma(E_{8}) =1$ and a trivial homomorphism $\phi$.
\end{itemize}
\begin{table}
\begin{eqnarray*} 
\begin{array}{|c|l|c|c|c|c|c|} \hline
  \rule{0mm}{5mm} w & K(w) & {\cal D}(w) &\overline\D (w) &  
  {\cal G} & \Gamma({\cal G})
  \\ \hline 
  \rule{0mm}{5mm} {\bf E_6}& -(ST)^{-1}& \{ST\}=\bbbz_{6}& \bbbz_2&
  E_6& \bbbz_2
  \\ \hline
  \rule{0mm}{5mm} {\bf E_7}& ~~~S & \{S\}=\bbbz_4& 1 & E_7 & 1 
  \\ \hline
  \rule{0mm}{5mm} {\bf E_8}& ~~~ST &\{ST\}= \bbbz_6 & 1 & E_8& 1 
  \\ \hline
\end{array}
\end{eqnarray*} 
\caption{Duality groups and graph automorphisms for ${\bf E_{N}}$ 
configurations.}  
 \end{table}

\section{Duality groups for affine configurations}
\setcounter{equation}{0}
In this section we will try to extend the result of previous section
to the case of affine exceptional configurations
$\widehat{\widetilde{\bf E}}_0$, $\widehat{\widetilde{\bf E}}_1$, and
the series $\widehat{{\bf E}}_N$ for $ 1\leq N \leq 8$. These 
configurations are more interesting because of their relation with del
Pezzo surfaces \cite{HI} but at the same time more difficult to analyze since
for these configurations ${\cal H}(w)\neq {\cal H}_{W}(w)$ i.e, not
all transpositions are of Weyl type.  This means that there 
are elements in ${\cal H}(w)$ whose action on roots may be
outer automorphisms of the root lattice.  Our strategy in this case
will be to find the map  $\overline{{\cal D}}(w)/\{-\unit\}\mapsto \pm\Gamma$ for
a fixed set of transformations 
used to undo the $\sl2z$ transformations
in ${\cal D}(w)$.

We begin with some general remarks applicable to the the affine
exceptional brane configurations ${\bf \widehat{E}_{N}}={\bf
  A^{N-1}BCBC}$ with $2\leq N \leq 8$.  All such configurations have
at least one ${\bf A}$ brane.  They have monodromy $K({\bf
  \widehat{E}_{N}})=T^{ 9-N }$, and one readily finds that Stab$(K) =
\{ -\unit, T \}$.  We now show that $T \in \D$ for $n>1$ by an
explicit calculation quite similar to that given in \myref{trdn}.  We
make the ${\bf A}$ brane do a counterclockwise 
round trip around the other branes:
\begin{eqnarray}
  {\bf ABCBC}
  \transform{T} 
  {\bf A X_{[0,-1]} X_{[2,1]}X_{[0,-1]}X_{[2,1]}}  
  \hskip-2pt\transform{
    P_1P_2P_3P_4P_4P_3P_2P_ 1} 
  {\bf ABCBC}\,.
\label{xxx}
\end{eqnarray}
Let $\{Q_{A},Q^{1}_{B},Q^{1}_{C},Q^{2}_{B},Q^{2}_{C}\}$ denote the
invariant charges on the branes. The transformed invariant charges are
found to be
\begin{eqnarray} \nn
  Q_{A}&\transform{T}& Q_{A}+q\,,\\ \nn 
  Q^{1}_{B}&\transform{T}& Q^{1}_{B}+Q_{A}\, ,\\  
  Q^{1}_{C}&\transform{T}&Q^{1}_{C}+Q_{A}\,,\\  \nn
  Q^{2}_{B}&\transform{T}& Q^{2}_{B}-Q_{A}\,,\\ \nn
  Q^{2}_{C} &\transform{T}& Q^{2}_{C}-Q_{A}\,. \nn
\end{eqnarray}
Here, $q=Q^{1}_{C}+Q^{2}_{C}-Q^{1}_{B}-Q^{1}_{B}$ is the total
$q$-charge of the junction $\mJ$. Since we are only interested in
junctions that correspond to states in the root lattice, we set $q=0$.
We see that the effect of the $T$ transformation on a junction of
$\widehat{E}_{N}$ with zero $q$ charge is simply
\begin{equation}
  \label{doti}
  {\bf J} \mapsto {\bf J}+Q_A({\bf J})\,\,{\mathbold \delta}\,.
\end{equation}
Here, $Q_{A}(\mJ)$ is the invariant charge on the ${\bf A}$-brane used
to undo the effect of the $T$ transformation, and ${\mathbold \delta}
= {\bf b}_{1}+{\bf b}_{2}-{\bf c}_{1}-{\bf c}_{2}$ \cite{infinite}.
Indeed, when there is more than one ${\bf A}$ brane, any of them can
be used to undo the effect of $T$. We have therefore shown that
\begin{equation}
  \D({\bf
    \widehat{E}_N})=\{-\unit, T\}, \qquad 2 \leq N \leq 8 \,.
\end{equation}  
${\bf \widehat{E}_{9}}$ being the composition of two copies of $D_4$
has ${\cal D}({\bf \widehat{E}_9})=\sl2z$.  In addition, since $K({\bf
  \widehat{E}_{N}})=T^{ 9-N }$ we also have
\begin{equation}
  \overline \D ({\bf
    \widehat{E}_N})=\{-\unit, T \,| \,T^{9-N} = 
  1\} = \bbbz_{2} \times \bbbz_{9-N},
  \qquad 2 \leq N \leq 8 \,.
\end{equation}  

The group Aut$(Q)/W$ of an affine algebra, written as $\pm \Gamma$
in section 4, is more precisely written as
\begin{equation}
  \hbox{Aut} (Q) /W  =  \bbbz_{2} \times \Gamma 
\end{equation}
where the element $(-1,e)$, with $e$ the identity in $\Gamma$, is the
transformation $Q\to -Q$ reversing the sign of all the vectors in the
root lattice, and thus reversing the sign of all Dynkin labels.  An
element of the form $(0, h\in \Gamma)$ simply acts by the graph
automorphism $h$ of the Dynkin graph of the affine algebra. The
duality $-\unit\in {\cal D}({\bf \widehat{E}_N})$ maps to $(1, 0)$ in
$\overline \D$ and then
\begin{equation}
  \phi : (1, 0) \in \overline \D = \bbbz_{2} \times \bbbz_{9-N} \to
  (1,e) \in \bbbz_{2} \times \Gamma
\end{equation}
Our computations will require finding how $T\in \D$ acts. For this we
note that it maps to $(0,1) \in \overline\D$.
We will find that 
\begin{equation}
  \label{hoft}
  \phi : (0, 1) \in \overline \D = \bbbz_{2} \times \bbbz_{9-N} \to
  (0,h(T)) \in \bbbz_{2} \times \Gamma \,, 
\end{equation}
where $h(T)$ is a graph automorphism.   This map
respects the product structure of the groups involved.

To simplify the formulae we also introduce the following notation for
Weyl transformations,
\begin{equation}
  W_{i_{1}^{n_{1}}i_{2}^{n_2}\cdots i_{k}^{n_k}} \equiv
  W_{n_1\alpha_{i_1}+n_2\alpha_{i_2}+\cdots n_{k}\alpha_{i_k}}\,.
\end{equation}
Let us now consider in detail the various configurations in the above
series. We will show the brane configurations and indicate the simple
root junctions. Then we select an ${\bf A}$ brane to undo the $T$
duality and use equation \myref{doti} to find the action on the simple
roots. The final step is writing this action as the composition of a
Weyl transformation and the action arising from a Dynkin graph
automorphism.  The answer is the graph automorphism $h(T) \in \Gamma$
defined in equation
\myref{hoft}.

\begin{itemize}
\item {\bf ${\bf \widehat{E}_{2}}$:} The $T$ transformation acts
  trivially on the roots since the ${\bf A}$-brane supports no root.
  Therefore
  \begin{equation}
    h(T)=0\,.
  \end{equation}
  
  \onefigure{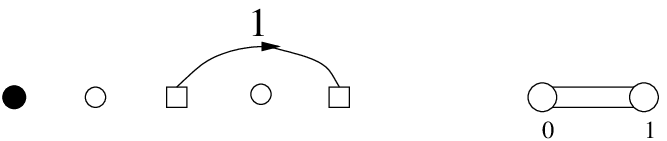}{$\widehat{{\bf E}}_2$ brane configuration and
    $\widehat{A}_{1}$ Dynkin diagram.}

\item {\bf ${\bf \widehat{E}_{3}}$:} Using the 
  rightmost ${\bf A}$ brane of the figure, $T$ acts on the simple
  roots as follows,
  \begin{eqnarray}
    T:(\alpha_0,\,\alpha_1,\,\alpha_2\,;\,\beta_0\,,\beta_1)
    &\rightarrow&
    (\alpha_0-\delta,\,\alpha_1+\delta,\,\alpha_2\,\, ;\,
    \,\beta_0 + \delta\,,\beta_1- \delta )\,.
    \label{T3} 
  \end{eqnarray} 
  Here, $(\alpha_{0}\, ,\alpha_{1}\,,\alpha_{2})$ are the roots of
  $\widehat{A}_{2}$ and $(\beta_{0}\,,\beta_{1})$ are the roots of
  $\widehat{A}_1$.  It is easy to verify that acting on simple roots
  \begin{equation}
    T = \Bigl(W_{\alpha_1}W_{\alpha_2} {\cal O} \,,   
              W_{\beta_0} \,           {\cal O}' \Bigr)\,,  
  \end{equation}
  where the first and second terms indicate the action on the
  $\widehat A_2$ and $\widehat A_1$ roots respectively. The graph
  automorphisms ${\cal O}$ and
  ${\cal O}'$ are  
  \begin{equation} 
    {\cal O}
    (\alpha_{0},\,\alpha_{1},\,\alpha_{2} )  =
    (\alpha_{2},\,\alpha_{0},\,\alpha_{1})\,,
    \qquad {\cal O}' (\beta_{0}\,,\beta_{1})
    =(\beta_{1}\,,\beta_{0})\,.
  \end{equation}
  These are elements of $\Gamma (
  \widehat{A}_{2}\oplus\widehat{A}_{1}) = {\cal D}_{6} \times
  \bbbz_{2} $, where $\D_6$ denotes the symmetry group of the
  triangle, here formed by the $\widehat A_2$ simple roots, with
  ${\cal O}$ the elementary rotation. In addition, ${\cal O}'$ is the
  nontrivial element of $\bbbz_{2}$, representing the exchange of the
  two simple roots of $\widehat A_1$. In summary;
  \begin{equation}
    h(T)=( {\cal O}\,, {\cal O}' ) \in {\cal
      D}_{6} \times \bbbz_{2} = \Gamma (
    \widehat{A}_{2}\oplus\widehat{A}_{1})\,. 
  \end{equation}
    
  \onefigure{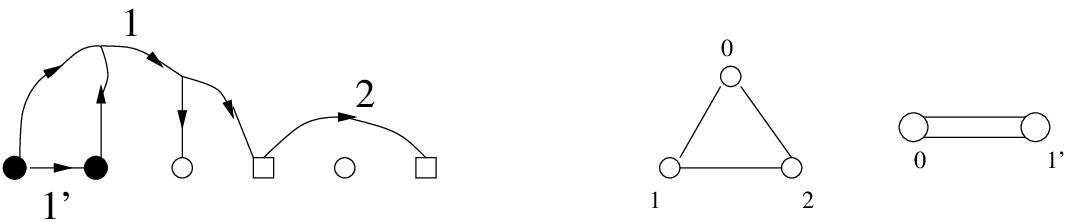}{$\widehat{{\bf E}}_{3}$ brane configuration and
    $\widehat{A}_{2}\oplus\widehat{A}_{1}$ Dynkin diagram.}

\item {\bf ${\bf \widehat{E}_4}$:}
  Using the rightmost ${\bf A}$ brane the nontrivial action of $T$ on
  the roots is given by
  \begin{equation}
    T : (\alpha_{2},
    \,\alpha_{4})
    \rightarrow (\alpha_{2}-\delta,\,
    \alpha_{4}+\delta)\,.
    \label{T4}
  \end{equation}
  A calculation shows that
  \begin{equation}
    T={\cal O} ~W_1W_2W_3W_0W_1W_2\,,\qquad
    {\cal O}(\alpha_{0},\,\alpha_{1},
    \,\alpha_{2},\,\alpha_{3},\,\alpha_{4})
    = (\alpha_{3},\, \alpha_{4},\,\alpha_{0},
    \,\alpha_{1},\,\alpha_{2})\,.
  \end{equation}
  Here ${\cal O}$ implements the transformation $\omega^3 \in
  \Gamma(\widehat A_4) = \D_{10}$, where $\omega$ is a cyclic minimal
  rotation of the pentagon.  Note that $\omega^3$ is a generator for
  the $\bbbz_{5}$ subgroup of rotations of $\D_{10}$.  Thus, in
  summary
  \begin{equation}
    h(T)={\cal O} = \omega^3 \in \D_{10}\,, 
  \end{equation}
  Note that in $\overline\D$, $T^5 = \unit$, and this is consistent
  with the map to $\D_{10}$.
  
  \onefigure{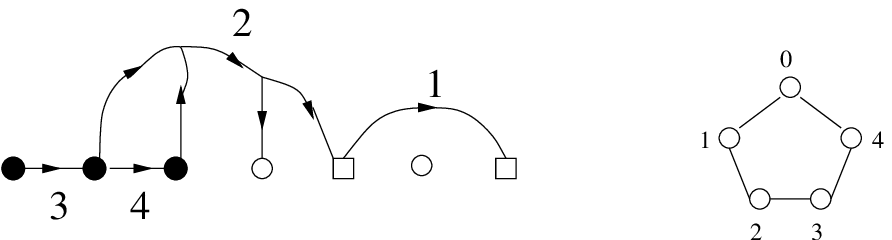}{$\widehat{{\bf E}}_4$ brane configuration and
    $\widehat{A_4}$ Dynkin diagram.}
  
\item {\bf ${\bf \widehat{E}_5}$:}
  Using the rightmost ${\bf A}$ brane, the action of $T$ is:
  \begin{equation}
    T :
    (\alpha_{0},\,\, \alpha_{5})
    \rightarrow (\alpha_{0}-\delta,\, 
    \alpha_{5}+\delta)\,, 
    \label{T5}
  \end{equation}
  This time we find
  \begin{equation}
    T={\cal O}\,W_2W_{1234}W_{35}\,,
    \qquad  {\cal O}(\alpha_{0},
    \,\alpha_{1},\,\alpha_{2},\,\alpha_{3},\,\alpha_{4},\,\alpha_{5})
    = (\alpha_{4},\,
    \alpha_{5},\,\alpha_{3},\,\alpha_{2},\,\alpha_{1},\,\alpha_{0})
  \end{equation}
  Here ${\cal O} \in \Gamma(\widehat D_{5}) = \D_8$, is a generator
  for the $\bbbz_{4}$ subgroup of $\D_8$, consistent with $T^4 =
  \unit$ in $\overline\D$. In summary:
  \begin{equation}
    h(T)={\cal O}\in \Gamma(\widehat D_{5}) = \D_8\,.
  \end{equation}
  \onefigure{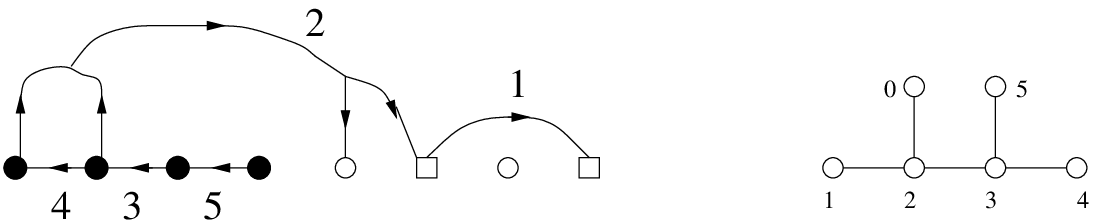}{$\widehat{{\bf E}}_5$ brane configuration and
    $\widehat D_{5}$ Dynkin diagram.}
  
\item {\bf ${\bf \widehat{E}_6}$:}
  Using the leftmost brane, one finds that $T$ induces the
  transformations
  \begin{equation}
    T:
    (\alpha_{0},\,\alpha_{5})\rightarrow
    (\alpha_{0}-\delta,\,\alpha_{5}+\delta\,)\,.
  \end{equation}
  A calculation shows that
  \begin{equation}
    T= {\cal O} ~
    W_{34}W_{1234}W_{1236}W_{23}W_{45}\,, 
    \label{T6}
  \end{equation}
  where ${\cal O}\in \Gamma(\widehat E_6)=S_{3}$ is the generator of
  the $\bbbz_{3}$ subgroup of $S_3$ performing the rigid minimal
  rotation of the Dynkin diagram:
  \begin{equation}
    {\cal O}(\alpha_{0},\,\alpha_{1},
    \,\alpha_{2},\,\alpha_{3},\,\alpha_{4},\,\alpha_{5},\,\alpha_{6})
    = (\alpha_{1},\,
    \alpha_{5},\,\alpha_{4},\,\alpha_{3},\,\alpha_{6},
    \,\alpha_{0},\,\alpha_{2})\,.
  \end{equation}
  This is compatible with $T^3 = \unit$ in $\overline \D$. In summary:
  \begin{equation}
    h(T)= {\cal O}\in \Gamma(\widehat E_6)=S_{3}\,.
  \end{equation}
  \onefigure{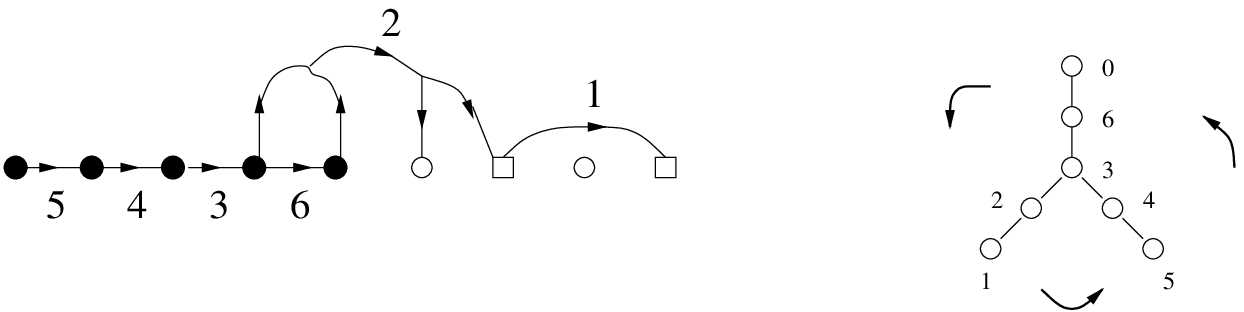}{$\widehat{{\bf E}}_6$ brane configuration and
    $\widehat E_6$ Dynkin diagram.}

\item {\bf ${\bf \widehat{E}_7}$:} 
  Using the leftmost ${\bf A}$ brane, $T$ acts as
  \begin{eqnarray}
    &T&:
    (\alpha_{0},\,\alpha_{1} )\rightarrow
    (\alpha_{0}-\delta,\,\alpha_{1}+\delta  )
    \nn \\
    &T&={\cal O}
    ~W_{45}W_{45^{2}67}W_{123^{2}4567}
    W_{12^{2}3^{2}4^{2}57}W_{34}\,,
    \label{T7}
  \end{eqnarray}
  where ${\cal O}\in \Gamma(\widehat E_7 )=\bbbz_{2}$ is the
  nontrivial generator of the graph automorphism, and exchanges the
  two long branches of the Dynkin diagram:
  \begin{equation}
    {\cal O}(\alpha_{0},\,\alpha_{1},
    \,\alpha_{2},\,\alpha_{3},\,\alpha_{4},\,\alpha_{5},
    \,\alpha_{6},\,\alpha_{7})= (\alpha_{1},\,
    \alpha_{0},\,\alpha_{6},\,\alpha_{5},\,\alpha_{4},
    \,\alpha_{3},\,\alpha_{2},\,\alpha_{7})\,.
  \end{equation}
  This is compatible with $T^2 = \unit$ in $\overline \D$.  In
  summary:
  \begin{equation}
    h(T)= {\cal O}\in \Gamma(\widehat E_7)=\bbbz_{2}\,.
  \end{equation} 

  \onefigure{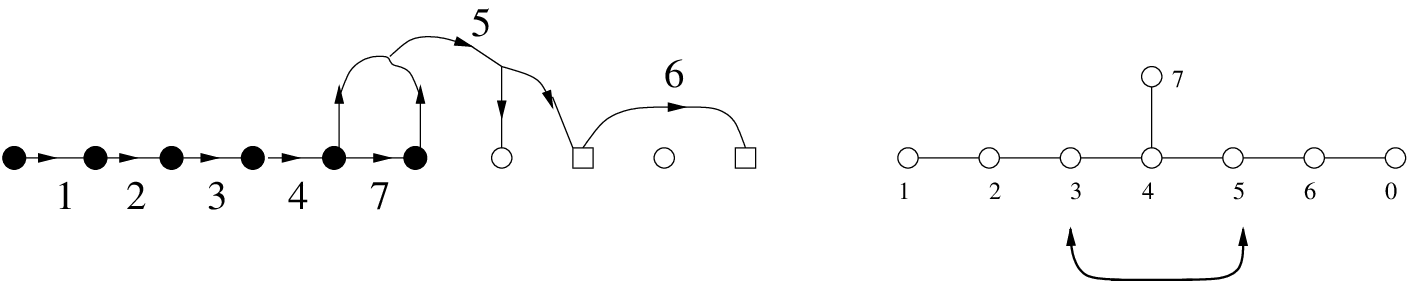}{$\widehat{{\bf E}}_7$ brane configuration and
    $\widehat E_7$ Dynkin diagram.}

\item {\bf ${\bf \widehat{E}_8}$:} 
  Using the leftmost ${\bf A}$ brane the $T$ action is:
  \begin{equation}
    T:(\alpha_{0},\alpha_{1} )\mapsto 
    (\alpha_{0}-2\delta,\alpha_{1}+\delta ) 
  \end{equation}
  Since $\Gamma(\widehat E_8)=1$, the above ought to be a pure Weyl
  transformation.  Indeed,
  \begin{eqnarray}
    T= W_0W_{123456}W_{123458}W_{34567}
    W_{2345^{2}678}W_{6}W_{345^{2}678}
    W_{234568} \nn\\  \cdot W_{56}
    W_{1234567}W_{12} 
    W_{12345678}W_{34568}
    W_{12345^{2}68}\,.   
  \end{eqnarray}
  Therefore $h(T) = e$ is the identity element in the trivial group
  $\Gamma(\widehat E_8) =1$. Indeed $T=\unit$ in $\D$ as well.
  \onefigure{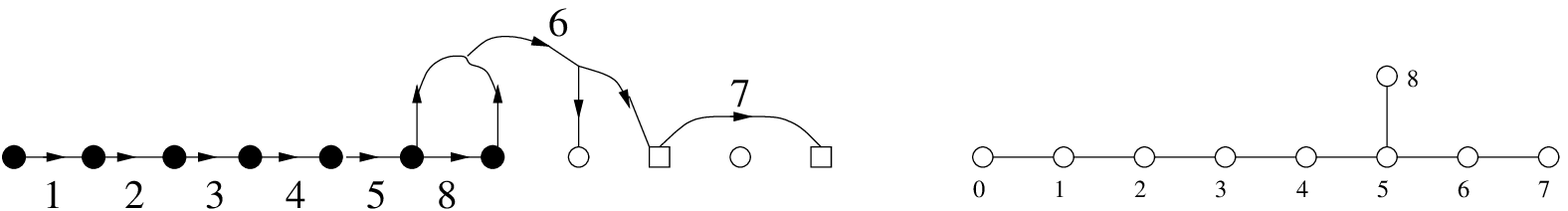}{$\widehat{{\bf E}}_8$ brane configuration and
    $\widehat E_8$ Dynkin diagram.}
\end{itemize}

We now consider in detail the special cases:

$\bullet$ ${\bf \widehat{\tilde{E}}_0}$: This configuration consists
of three 7-branes ${\bf X_{[2,-1]}X_{[-1,2]}X_{[1,1]}}$ (eq(3.10) of
\cite{infinite} with last two branes interchanged). Therefore
$K({\bf\widehat{E}_0})=T^{9}$ and $\mbox{Stab}(K)=\{-\unit, T\}$. An
important property of this configuration is that the asymptotic charge
$(p,q)$ of every junction satisfies the condition $p-q \equiv
0\,(\mbox{mod}\,3)$. An argument similar to the one used for
${\bf D_0}$ proves that if $T^{k}\in \D(w)$ then $k \equiv
0\,(\mbox{mod}\,3)$. By explicit computation one can show that
$T^{3}\in {\cal D}(w)$:
\begin{eqnarray}
\label{tre3}
  {\bf X_{[2,-1]}X_{[-1,2]}X_{[1,1]}} \transform{T^{3}} {\bf
    X_{[-1,-1]}X_{[5,2]}X_{[4,1]}}
  \hskip-2pt\transform{P_{2}^{-1}P_{1}^{-1}}  
  {\bf X_{[2,-1]}X_{[-1,2]}X_{[1,1]}}\,.
\end{eqnarray}
Thus $\D( {\bf \widehat{\tilde{E}}_0})=\{-\unit, T^{3}\}$ and
$\overline \D({\bf \widehat{\tilde{E}}_0})=\{-\unit,
T^{3}~|~T^{9}=\unit\} = \bbbz_{2}\times \bbbz_3$. This configuration
supports no algebra. The only junctions having no asymptotic charges
(localized) are multiples of the delta junction 
\begin{equation}
\label{tre33}
{\bf \delta}= {\bf x_{[2,-1]}}+{\bf x_{[-1,2]}}-{\bf x_{[1,1]}} \,, 
\end{equation}
which can be represented as a $(-1,0)$ string circling the 
branes in the counterclockwise direction.     
The action of the $T^3$ duality transformation on a general
junction is found as usual using the brane transpositions in 
\myref{tre3}. Denoting the invariant charges on the 
 ${\bf X_{[2,-1]}, X_{[-1,2]},\hbox{and} \, X_{[1,1]}}$ branes by
$Q_1, Q_2$ and $Q_3$ respectively,  we find that 
\begin{equation}
\label{map2}
T^3 :  (Q_1,
Q_2, Q_3) \mapsto (Q_2\,,\,  -Q_3\, ,\,  -Q_1 + 3 (Q_2 + Q_3)\, ) \,.
\end{equation} 
In this notation the delta junction is $(1,1,-1)$, and one readily
verifies that $T^3$ leaves it invariant. This was expected, since
$T^3$ leaves invariant the $(-1,0)$ string, and brane transpositions
cannot affect a looping string.

$\bullet$ ${\bf \widehat{\tilde{E}}_1}$: Here $K({\bf
\widehat{E}_1})=T^{8}$ and $\mbox{Stab}(K)=\{-\unit, T\}$. Since this
brane configuration has an ${\bf A}-$brane a simple computation shows
that $T \in\D(w)$:
\begin{equation}
  {\bf AX_{[2,-1]}X_{[-1,2]}X_{[1,1]}} \transform{T} {\bf
    AX_{[1,-1]}X_{[1,2]}X_{[2,1]}}
  \hskip-2pt\transform{P_{1}P_{2}P_{3}P_{3}P_{2}P_{1}}  
  {\bf AX_{[2,-1]}X_{[-1,2]}X_{[1,1]}}\,.
\end{equation}
Thus $\D({\bf \widehat{\tilde{E}}_1})=\{-\unit, T\}$ and $\overline
\D({\bf \widehat{\tilde{E}}_1})=\{-\unit, T~|~T^{8}=\unit\} =
\bbbz_{2}\times \bbbz_8$. A  localized junction on this configuration is
specified by  the number of ${\bf \delta}$ loops (the junction
 in \myref{tre33}) 
and a $u(1)$ charge $\bar Q$. We thus write 
$\mJ= 
m{\bf \delta}+ \bar Q\overline{\mJ}$ \cite{SZ} where 
\begin{equation}
\label{thatj}
{\bf \overline{J}}=3{\bf a}-{\bf x_{[2,-1]}}-{\bf x_{[1,1]}}.
\end{equation}
Under the $T$ 
transformation the junction $\mJ\equiv ( m, Q)$ transforms as
\begin{equation}
T: (m,\,\bar Q) \mapsto
(\,m+3\bar Q\, , \,\bar Q ) \,, 
\end{equation}
which follows because the ${\bf A}$ brane that must be circled
around has invariant charge $3\bar Q$. 

Since the monodromy of the configuration
is $T^8$, we see that 
\begin{equation}
\label{isit}
T^8: (m,\,\bar Q) \mapsto
(\,m+24\bar Q\, , \,\bar Q ) \,, 
\end{equation}
should be a transformation that can be generated simply
by crossing transformations.  Such a tranposition 
(actually its inverse) was discussed in
section 3.2. 
It is not of Weyl type because there are no real root
 junctions for this brane
configuration.

Indeed, the required  transposition  is the inverse of 
that discussed in section 3.2 for
the  case of $\hat{\tilde{{\bf E}}}_1$.  These 
transpositions are equivalent to first taking the ${\bf A}$ brane {\it clockwise}
around
the other three
 branes three times. This has the effect of changing
the
 labels of the other three branes as if acted by $T^3$. 
 Those
branes are then restored to their original labels
 by performing the
transpositions indicated in the 
 second step of \myref{tre3}. The
first step takes 
 a junction $\mJ = m{\bf \delta} + \bar Q
\overline\mJ$
 and adds to it $(-3 Q_A) = -9\bar Q$ delta junctions. 
This step, while changing the labels of the three rightmost
 branes,
it does not change the invariant charges they
 have; these are, in the
notation used for ${\bf \widehat{\tilde{E}}_0}$,
 $\bar Q (-1, 0, -1)$
(see \myref{thatj}). Using \myref{map2}
 we see that under the
restoring transposition: 
 $\bar Q (-1, 0, -1) \mapsto \bar Q (0, 1,
-2) = \bar Q (-1, 0, -1) 
 + \bar Q {\bf \delta}$.  Therefore, the
complete series
 of transpositions $\tilde b$ adds $(-9 +1 ) \bar Q =
-8 \bar Q$ delta junctions:
\begin{equation}
\tilde b : (m, \bar Q) \mapsto  (m - 8\bar Q, \bar Q)
\end{equation}
Comparing with \myref{isit} we see that indeed $T^8$ has
the same effect as the transposition $(\tilde b)^{-3}$.
Since a $T$ duality adds $3\bar Q$ delta junctions, 
and we can add or remove $8\bar Q$ delta junctions by
transpositions, $T^8$ is the lowest power of $T$ that
is equivalent to a transposition.

$\bullet$ ${\bf \widehat{E}_1}$: Here $K({\bf \widehat{E}_1})=T^{8}$.
A junction of asymptotic charge $(p,q)$ with support on this
configuration satisfies the condition $p+q\equiv 0\,(\mbox{mod}\,2)$.
Thus if $T^{k}\in \D(w)$ then $k \equiv 0\,(\mbox{mod}\,2)$. By
explicit computation one can show that $T^{2}\in \D(w)$:
\begin{equation}
  {\bf BCBC} \transform{T^{2}} {\bf
    X_{[-1,-1]}X_{[3,1]}X_{[-1,-1]}X_{3,1]}}
  \hskip-2pt\transform{P_{3}^{-1}P_{1}^{-1}}  
  {\bf BCBC}\,.
\end{equation}
Thus $\D( {\bf \widehat{E}_1})=\{-\unit, T^{2}\}$ and $\overline
\D({\bf \widehat{E}_1})=\{-\unit, T^{2}~|~T^{8}=\unit\} =
\bbbz_{2}\times \bbbz_4$.  The $T^{2}$ transformation acts on the
invariant charges in the following way,
\begin{equation}
  T^{2}: (Q_{B_{1}}, Q_{C_{1}}, Q_{B_{2}}, Q_{C_{2}}) \mapsto
  (Q_{C_{1}}, -Q_{B_{1}}+2Q_{C_{1}}, Q_{C_{2}}, -Q_{B_{2}}+2Q_{C_{2}})\,.
  \label{inv}
\end{equation}
The invariant charges of the root junctions and the delta junction
${\bf \delta}$ are \cite{infinite}
\begin{equation}
{\bf { \malpha_0}} = (1,0,-1,0)\,\,,\,\,{\bf {
\malpha_1}}=(0,1,0,-1)\,\,,\,\,{\bf {\bf \delta}}={ \malpha_0}+{
\malpha_1}=(1,1,-1,-1)\,.
\label{invroots}
\end{equation}
Thus from (\ref{inv}) and (\ref{invroots}) it follows that
\begin{equation}
  T^{2}: (\alpha_{0}\,,\alpha_{1}) \rightarrow
  (\alpha_{0}-\delta\,,\alpha_{1}+\delta)\,.
\end{equation}
$\Gamma(\widehat A_1)=\bbbz_{2}$ where the nontrivial element
 is 
${\cal O}(\alpha_{0}\,,\alpha_{1}) = 
(\alpha_{1}\,,\alpha_{0})$. One can verify that acting on the simple roots 
\begin{equation}
  T^{2}={{\cal O}}W_{0}\,.
\end{equation}
Therefore, $\phi:\overline \D({\bf \widehat A_1})\mapsto
\bbbz_{2}\times \Gamma(\widehat A_1)$, is determined by
\begin{equation}
  h(T^{2})={{\cal O}}\,.  
\end{equation}
This completes our analysis of the duality groups of
affine exceptional brane configurations.

\onefigure{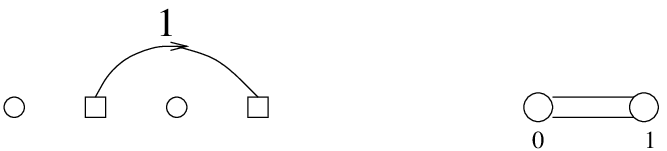}{${\bf \widehat{E}_1}$ brane configuration and
$\widehat{A}_1$ Dynkin diagram.}  
\begin{table}
\begin{eqnarray*}   
\begin{array}{|c|l|c|c|c|c|c|} 
  \hline\rule{0mm}{5mm} 
  w & K(w) & {\cal D}(w) 
  &\overline{\cal D}(w)& {\cal G} & 
  \hbox{Aut} (Q) /W
  \\ \hline\rule{0mm}{6mm} 
  {\bf \widehat{\widetilde E}_{0}}    & T^{9}      
  &  \{-\unit, T^{3}\} &  \bbbz_{2}\times \bbbz_{3}        
  & -- &  --   \\ \hline
\rule{0mm}{6mm} 
 {\bf \widehat{E}_{1}}    & T^{8}       
  &  \{-\unit, T^{2}\} &  \bbbz_{2}\times \bbbz_{4}          
  & \widehat{A}_{1}  &  \bbbz_{2}\times \bbbz_{2}      
 \\ \hline
\rule{0mm}{6mm} 
{\bf \widehat{\tilde{E}}_{1}}    & T^{8}       
  & \{-\unit, T\} & \bbbz_{2}\times \bbbz_{8} & \widehat{u(1)}& --      
 \\ \hline
\rule{0mm}{5mm} 
  {\bf \widehat{E}_{2}}    &      T^{7}  
  & \{-\unit, T\} &  \bbbz_{2}\times\bbbz_{7}          
  &    \widehat{A}_{1}\oplus \widehat{u(1})                &
\bbbz_{2}\times   
\bbbz_{2}     
 \\ \hline\rule{0mm}{5mm} 
  {\bf \widehat{E}_{3}}    &    T^{6}    & \{-\unit, T\}       
  & \bbbz_{2}\times\bbbz_{6}          
  &   \widehat{A}_{2} \oplus \widehat{A}_{1}& \bbbz_{2} \times ({\cal
D}_{6}\times
\bbbz_{2})  
 \\ \hline\rule{0mm}{5mm} 
  {\bf \widehat{E}_{4}}    &       T^{5} & \{-\unit, T\}       
  &  \bbbz_{2}\times \bbbz_{5}         
  &  \widehat{A}_{4}   & \bbbz_{2}\times {\cal D}_{10}       
 \\ \hline\rule{0mm}{5mm} 
  {\bf \widehat{E}_{5}} &  T^{4}  &  \{-\unit, T\}  & 
  \bbbz_{2}\times\bbbz_{4}  & \widehat{D}_{5}  & \bbbz_{2}\times  {\cal D}_8
 \\ \hline\rule{0mm}{5mm} 
{\bf \widehat{E}_{6}} & T^{3}  &  
  \{-\unit, T\}  & \bbbz_{2}\times\bbbz_{3}  & \widehat{E}_{6} 
  & \bbbz_{2}\times {\cal D}_6  
 \\ \hline\rule{0mm}{5mm} 
  {\bf \widehat{E}_{7}} &  T^{2} &  \{-\unit, T\}  &  
  \bbbz_{2}\times\bbbz_{2}  &  \widehat{E}_{7} & \bbbz_{2}\times \bbbz_2
 \\ \hline\rule{0mm}{5mm} 
  {\bf \widehat{E}_{8}} &  T  & \{-\unit, T\}  & \bbbz_{2}  &
  \widehat{E}_{8}  &  \bbbz_{2} 
\\ \hline\rule{0mm}{5mm} 
  {\bf \widehat{E}_{9}} &  \unit  &\sl2z  & P\sl2z  &
  \widehat{E}_{9}  &  ? 
  \\ \hline
\end{array}
\end{eqnarray*} 
\caption{Duality groups and graph automorphisms of ${\bf  
\widehat{E}_{N}}$
configurations. $\D_6= S_3$ is the
  permutation group of three objects and ${\cal D}_{2N}$,  the
  dihedral group, is the group of symmetries of
  the regular $n-$gon. } 
\end{table}
\section*{Acknowledgements}
We wish to acknowledge useful conversations with Y. Imamura and V.
Kac.

This research was supported in part by the US Department of Energy
under contract \#DE-FC02-94ER40818.

\end{document}